\documentclass[fleqn,usenatbib]{mnras}

\usepackage{newtxtext,newtxmath}

\usepackage[T1]{fontenc}

\DeclareRobustCommand{\VAN}[3]{#2}
\let\VANthebibliography\thebibliography
\def\thebibliography{\DeclareRobustCommand{\VAN}[3]{##3}\VANthebibliography}

\usepackage{graphicx}	
\usepackage{amsmath}	

\title[IISM and orbital analysis of PSR~J1909$-$3744]{Analysis of the ionized interstellar medium and orbital dynamics of \\ PSR~J1909$-$3744 using scintillation arcs}

\author[Askew et al.]{
J.~Askew,$^{1,2}$ 
D.~J.~Reardon,$^{1,2}$
and R.~M.~Shannon$^{1,2}$
\\
$^{1}$Centre for Astrophysics and Supercomputing, Swinburne University of Technology, P.O. Box 218, Hawthorn, Victoria 3122, Australia\\
$^{2}$Australian Research Council Centre of Excellence for Gravitational Wave Discovery (OzGrav)\\
}

\date{Accepted 2022 October 21. Received 2022 October 19; in original form 2022 September 12}

\pubyear{2022}

\begin{document}
\label{firstpage}
\pagerange{\pageref{firstpage}--\pageref{lastpage}}
\maketitle

\begin{abstract}

Long-term studies of binary millisecond pulsars (MSPs) provide precise tests of strong-field gravity and can be used to measure neutron-star masses.
PSR~J1909$-$3744, a binary MSP has been the subject of several pulsar timing analyses.
The edge-on orbit enables measurement of its mass using the Shapiro delay; however, there is degeneracy in the sense of the inclination angle, $i$, and multiple solutions for the longitude of ascending node, $\Omega$.
Radio pulsars scintillate due to inhomogeneities in the ionized interstellar medium (IISM).
This can result in scintillation arcs in the power spectrum of the dynamic spectrum that can use these to study the interstellar medium and constrain binary pulsar orbits.
Here, we study the scintillation of PSR~J1909-3744 using observations from the 64-m Parkes Radio Telescope (Murriyang) over $\approx$13\, years, using techniques to study scintillation in a lower signal-to-noise regime.
By monitoring annual and orbital variations of the arc-curvature measurements we are able to characterise the velocity of the IISM.
We find that the statistics of the IISM remained stationary over this time and a slightly anisotropic model (axial ratio $\gtrsim1.2$) is preferred.
We measure the relative distance to a single dominant thin scattering screen at $s=0.49\pm0.04$, or $D_s=590\pm50$\,pc, with an angle of anisotropy $\zeta=85\pm6^\circ$ (East of North) and velocity in the direction of anisotropy $V_{\textrm{IISM}, \zeta}=14\pm10$\,km\,s$^{-1}$.
By combining a physical model of the IISM and current pulsar timing results, we also constrain $\Omega=225\pm3^\circ$ and $i=86.46\pm0.05^\circ$.

\end{abstract}




\section{Introduction}\label{chapt:Introduction}

The study of interstellar scintillation using pulsars enables the characterisation of the ionized interstellar medium (IISM) and binary pulsars \citep{Lyne1982, Rickett1990, Bhat1998, Stinebring2001, Ord2002}.
Historically, investigation of precision orbital dynamics of pulsars has been done using pulsar timing \citep{Ruderman1975, Lattimer1990, Wolszczan1992, Kramer2006, Demorest2010}.
However, pulsar timing is only sensitive to radial motion \citep{Ransom2004}.
The orbital inclination angle, $i$, is needed to create a complete three-dimensional (3D) model of a binary system.
It is possible to measure the sine of the inclination angle from pulsar timing if the system is nearly edge on.
However, it is rare to be able to measure the sense of the inclination angle using this technique.
Scintillation, which is sensitive to transverse velocities, offers an alternative perspective when solving binary pulsar orbits.
By combining pulsar timing with scintillation models, we can create a full 3D model of the orbit.
This is applicable in particular for $i$ or the longitude of ascending node $\Omega$, which are difficult to determine using pulsar timing \citep{Rickett2014, Liu2020}.

A pulsar emits radio waves that are dispersed, refracted, and diffracted by inhomogeneities in the IISM \citep{Rickett1969, Coles1987, Rickett1990, Cordes2002}.
This can be manifested as interstellar scintillation, which are flux variations with time and frequency \citep{Armstrong1995}.
Islands of intensity, known as `scintles', appear in the dynamic spectrum of frequency against time for each observation \citep{Backer1975}.
While often dynamic spectra are defined to be the intensity as a function of time and frequency ($S(t,\,f)$), we will define it to be a function of time and wavelength ($S(t,\,\lambda)$) to remove the frequency dependence of arc curvatures. 


\begin{figure*}
	\includegraphics[width=1\columnwidth]{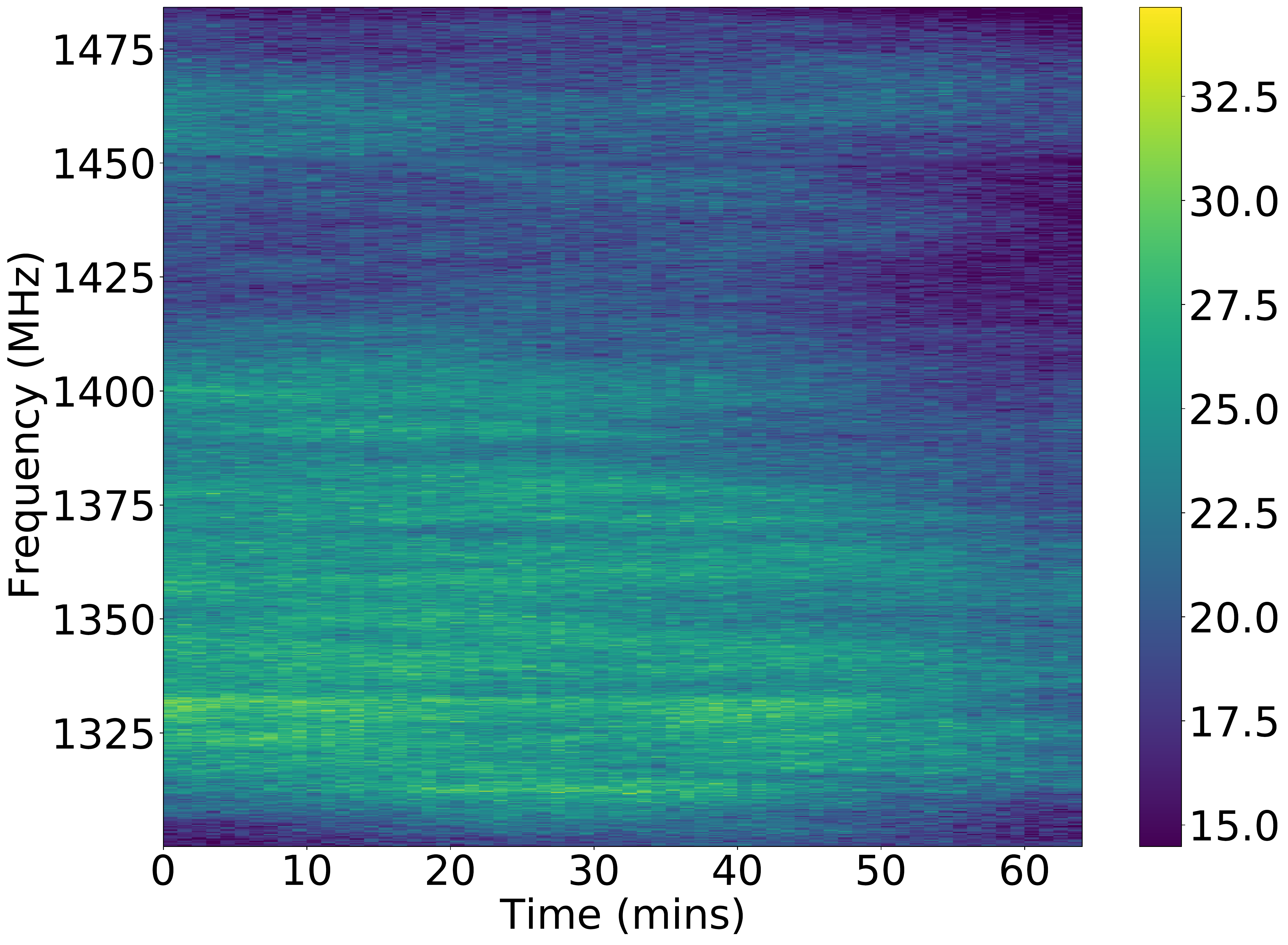}
	\includegraphics[width=1\columnwidth]{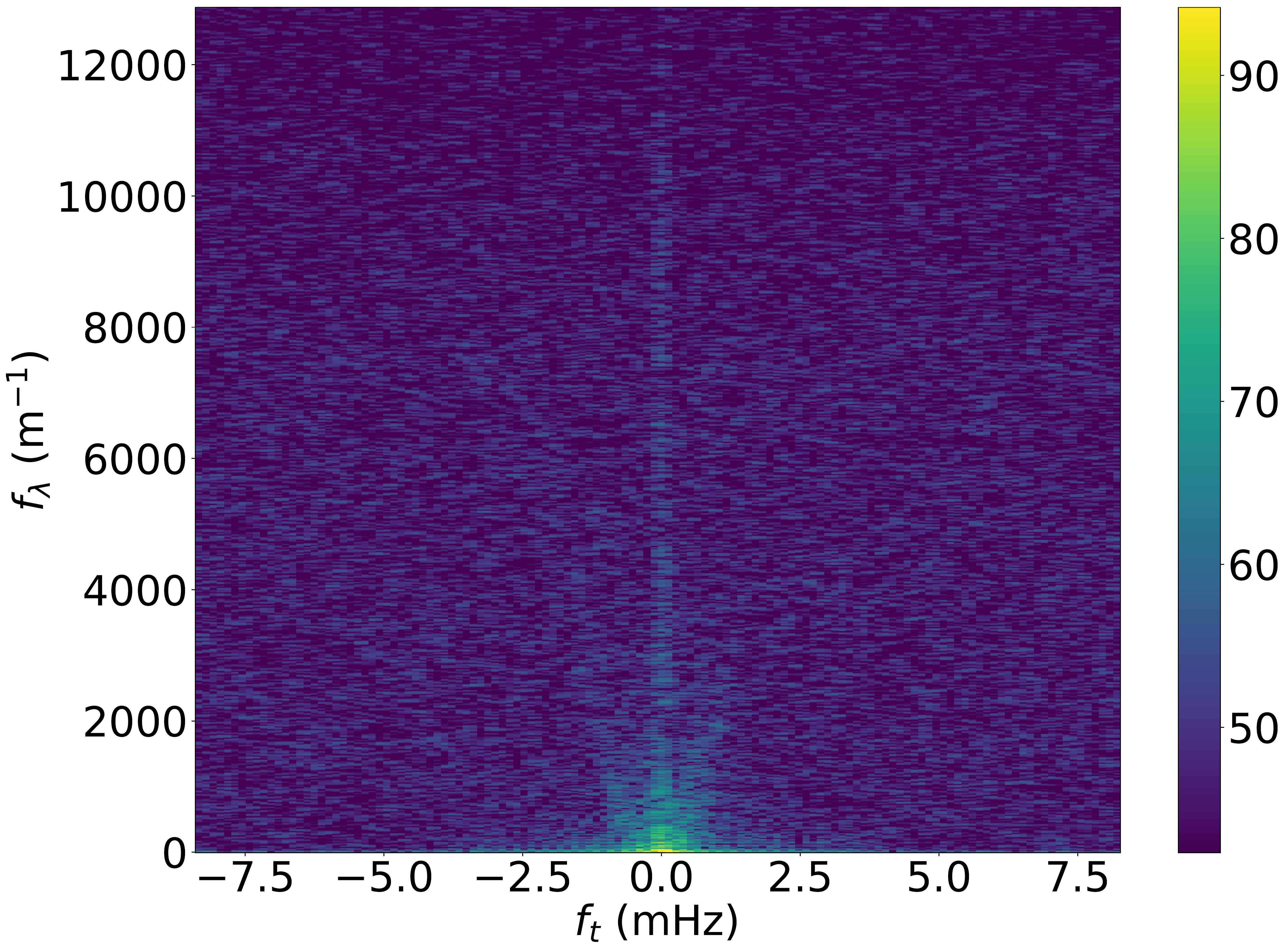}
    \caption{ Dynamic (left) and secondary spectra (right) of the highest $S/N$ scintillation arc. For this observation taken on 22/05/2011, the dynamic spectrum shows variation in the flux density (in mJy) received at the telescope at a given frequency and time. The secondary spectrum (in dB) demonstrates a scintillation arc apparent to the eye as well as noise along the $f_t=0$ axis. This noise is the result of correlated noise in frequency caused by RFI in the dynamic spectrum. We can also identify it as it has a different extent in Doppler ($f_\lambda$) compared to the main arc.}
    \label{fig:fig1}
\end{figure*}


The power spectrum of the dynamic spectrum, known as the `secondary spectrum', can show parabolic arcs of power \citep{Stinebring2001}.
When two points in the scattered image from angles $\boldsymbol{\theta}_1$ and $\boldsymbol{\theta}_2$ interfere, power in the secondary spectrum is manifested at a point ($f_t$,\,$f_\lambda$) where $f_t$ and $f_\lambda$ are conjugate variables to $t$ (time) and $\lambda$ (wavelength).
These are given by
\begin{eqnarray}
    f_t &&= \frac{1}{s\lambda_c}\boldsymbol{V}_{\textrm{eff}} \cdot \left(\boldsymbol{\theta}_2 - \boldsymbol{\theta}_1\right),
    \label{eqn:Doppler}
\end{eqnarray}
\begin{eqnarray}
    f_{\lambda} &&= \frac{D(1 - s)}{2s\,\lambda_c}\left(\boldsymbol{\theta}_2^2 - \boldsymbol{\theta}_1^2\right),
    \label{eqn:timedelay}
\end{eqnarray}
where $\lambda_c$ is the wavelength corresponding to the central frequency of the observation, $D$ is the distance to the pulsar, $\boldsymbol{V}_{\textrm{eff}}$ is the effective velocity along the line of sight, and $s = 1 - D_s/D$ is the fractional distance to the scattering screen, $D_s$ is the distance to the scattering screen from the observer.
As $f_t$ scales linearly with the scattering angles and $f_\lambda$ scales quadratically, there is a parabolic relation between $f_t$ and $f_\lambda$ where the parabolic curvature $\eta$ is defined as $f_\lambda = \eta f_t^2$.
Measurements of the curvature of these `arcs', $\eta$, can be used to measure properties of the IISM and $\boldsymbol{V}_{\textrm{eff}}$ \citep{Walker2004, Cordes2006}, through,
\begin{eqnarray}
    \eta = \frac{Ds\left(1 - s\right)}{2V_{\textrm{eff}}^2\cos^2{\psi}},
    \label{eqn:eta}
\end{eqnarray}
where $\psi$ is the angle between the major axis of anisotropy, and the effective velocity.
This angle, $\psi$, can be calculated from $\zeta$, the angle of anisotropy in terms of the spatial structure in Celestial coordinates (East of North).
In the case of the strong scattering regime with an isotropic screen, such that $\cos{\psi}\equiv1$, a sharp outer edge to the arc is seen \citep{Cordes2006}.

The velocity of the Earth, IISM, and pulsar motion (proper motion and orbit) affect $\boldsymbol{V}_{\textrm{eff}}$ \citep{Rickett2014}.
Long-term observations can be used to measure $\eta$ across the time scale of years and with multiple samples across orbital phase of a binary pulsar.
An observing campaign such as this can be used to measure variations in $\eta$ with annual and binary orbital phase.
These long-term analyses provide an accurate method to measure scintillation parameters \citep{Reardon2020} when compared to single epoch studies \citep[e.g.,][]{Bhat2016}.
The ability to detect arcs depends on the choice of observing frequency and the choice of pulsar.
Pulsars are brighter at lower frequencies, however, the structures in the scintillation pattern are also frequency dependent ($\propto f^{-4}$).
At low frequencies, the scintillation pattern can therefore become unresolved for most pulsar timing observing systems which have modest (few hundred kHz) resolution\footnote{It is possible to record data with higher frequency resolution. This can result in lower pulse phase resolution which is suboptimal for pulsar timing.}.

A millisecond pulsar (MSP) that is optimal for such studies is PSR~J1909$-$3744.
It is regularly observed at the Parkes radio telescope (Murriyang) with a flux density of $S_{1400}=2.5\pm0.2$\,mJy (at 1400\,MHz observing frequency) \citep{Dai2015}.
This pulsar is also timed to high precision with an rms timing residual of $\approx$99\,ns due to its narrow pulse width of $43\,\mu$s \citep{Jacoby2003, Reardon2021}.
Through pulsar timing, it is possible to measure the distance to the pulsar to be $D=1.152\pm0.003$\,kpc \citep{Reardon2021}.
The main purpose of the observations is to use them to search for nanohertz-frequency gravitational waves \citep{Shannon2015, Goncharov2021}, as part of the Parkes Pulsar Timing Array \citep[PPTA, ][]{Manchester2013}.
However, the regular observations over many years also make these data ideal for long-term scintillation studies.

In this paper, we model the long-term variations in arc curvature for PSR~J1909$-$3744.
We have achieved this within a low signal-to-noise ($S/N$) regime.
Ultimately, this study has led to a greater understanding of the characteristics of the IISM.
In Section \ref{chapt:Methods}, we describe our data processing as well as the methods and models considered.
The foundational results and outcomes of initial modelling are explored in Section \ref{chapt:Results}.
We find modest disagreement between models, and with published timing results.
In Section \ref{chapt:Discussion} we critically compare the models and investigate reasons for the discrepancies, including $\boldsymbol{V}_{\textrm{IISM}}$ alternatives.
This includes using physical priors on the IISM velocity and degree of anisotropy.
In Section \ref{chapt:Conclusions}, we summarise our findings.

\section{Methods}\label{chapt:Methods}

The methods presented here build on previous works \citep{Reardon2020, Walker2022}.
Below we explore some of these techniques and describe extensions that are critical to our analysis.

\subsection{Observations}\label{chapt:Observations}

Our studies focus on PSR~J1909$-$3744, an MSP observed as a part of the PPTA using Murriyang.
PSR J1909$-$3744 is in the most nearly circular orbit measured, with an eccentricity of $1.069\times10^{-7}$ and an orbital period of $\approx$1.533\,days.
These observations occur at a cadence of one epoch every three weeks, on average, with potentially multiple observations per epoch.
Previous works using PPTA data have included many studies of the IISM \citep{You2007, Keith2013, Coles2015, Reardon2020, Walker2022}.
Here we use data published with the PPTA second data release \citep[PPTA DR2; ][]{Kerr2020}.
Observations in PPTA DR2 cover four observing bands $40/50$\,cm (at centre frequencies, $\nu_c$ $\approx$685\,MHz and $\nu_c$ $\approx$732\,MHz respectively), $20$\,cm ($\nu_c$ $\approx$1369\,MHz), and $10$\,cm ($\nu_c$ $\approx$3100\,MHz).
The observations used in this analysis are from the $20$\,cm band, which was the only band where we detected scintillation arcs. 
Typical observations were $\sim$1\,hour in duration; observations shorter than 20\,min were not considered.
Our initial data set spans $\approx$15\,years, over which over $\approx$\,4000 observations have been collected at Murriyang. 
From these, 57 high $S/N$ scintillation arcs (resolved and apparent to the eye) were found between MJD 53728 to 58596 (December 2005 to April 2019, $\approx$13\,years). 
The small number of scintillation arcs found within this large dataset can be attributed to a few factors.
There were no arcs in the 10/50\,cm observing bands because of lower $S/N$, which included $\approx$40\% of the 4000 observations.
Some of the observations remained corrupted by radio frequency interference (RFI) induced noise close to the $f_t=0$ axis in the secondary spectrum.
As discussed in the appendix, we find with probability p = 0.96 bias against detecting high curvature arcs (see Figure \ref{fig:KolmogorovSmirnov}).

For our observations, it was determined that J1909$-$3744 was observed in the strong scattering regime, by estimating the Born variance as $m_b^2=0.773(\nu_c/\Delta\nu_d)^{5/6}$, where we took $\Delta\nu_d$ as the median scintillation bandwidth across all 57 observations and found $m_b^2\approx32$ \citep{Rickett1990, Reardon2020}.
Further information on the telescope systems and operation are found within \citet{Manchester2005} and \citet{Manchester2013} while the data processing and timing results can be found in \citet{Kerr2020} and \citet{Reardon2021} respectively.
 
The main data product for scintillation analysis is the dynamic spectrum, $S\left(t,\nu\right)$, (Figure \ref{fig:fig1}).
For our observations, the typical sub-integration time is ~$\approx$60\,s and channel bandwidth is ~$\approx$0.25\,MHz.
The data needed to produce dynamic spectra for PSR~J1909$-$3744 were pre-processed within the PPTA-DR2 data processing pipeline, which is based on \texttt{PSRCHIVE} \citep{Hotan2004}. 
The bandwidth of an observation varies as we excise channels affected by RFI resulting in bandwidths centered at $\approx$1369\,MHz ranging between $\approx$200$-$300\,MHz.
The dynamic spectrum for each observation was calculated using \texttt{psrflux} in the PPTA-DR2 pipeline and analysed using \texttt{scintools}\footnote{\url{https://github.com/danielreardon/scintools}} \citep{Reardon2020}.

To produce a secondary spectrum we follow the methods described in \citet{Reardon2020} and \citet{Walker2022}.
We begin by re-sampling the dynamic spectrum uniformly in wavelength, $S\left(t,f\right) \xrightarrow{} S\left(t,\lambda\right)$.
This removes the frequency dependence of $\eta$ as seen in Equations \ref{eqn:Doppler} and \ref{eqn:timedelay}.
It also has the effect of improving the sharpness of the arcs as they appear in the secondary spectrum.
Other methods re-sample temporally or decompose the power of the secondary spectrum relative to the two scattering angles $\theta_1$ and $\theta_2$ \citep{Sprenger2021}.
These methods would be useful for measuring the sub-structure of arcs, including inverted arclets.
However, we do not observe such substructure so choose a more computationally efficient procedure.
In addition to subtracting the mean flux, we apply a Hann window on the outer 10\% of the dynamic spectrum.
Finally, we perform the two-dimensional (2D) fast Fourier transform with zero-padding and take its squared magnitude which produces the associated power spectrum $P\left(f_t,f_\lambda\right)$. 
For some observations, it was also necessary to apply first-difference pre-whitening, followed by the appropriate post-darkening after Fourier transforming as described in \citet{Coles2011}.
This has the effect of reducing scattered power along the axes of the secondary spectrum.
This results in the secondary spectrum seen in Figure \ref{fig:fig1} (right panel) and is defined to be $P\left(f_t,\,f_\lambda\right) = 10\textrm{log}_{10}( | \Tilde{S}( t,\,\lambda) |^2 )$, where $\Tilde{S}( t,\lambda)$ is the Fourier transform of the mean-subtracted and windowed dynamic spectrum.

Before measuring $\eta$, a careful analysis of the noise in the secondary spectrum was undertaken.
Many observations had noise flagged close to the $f_t$, $f_\lambda$=0 axes.
The power interior to the arc, along the $f_t$ axis, can have a few origins.
It could be a bonafide secondary arc (which we were not sensitive to), noise caused by RFI, or increased power interior to the arc due to anisotropic scattering (when the scattering strength is strong). 
For each observation, the $S/N$ ratio of the arc was determined.
We define the on-arc region as one standard deviation in curvature away from the peak power, and we define the off-arc region as $|f_t| > 2$\,mHz.
The $S/N$ ratio is defined as the ratio of the weighted summed power of the on-arc to the off-arc, with the power spectrum along $f_\lambda$ providing the weights \citep{Reardon2020}.

The 57 arcs used were the brightest, with $S/N\geq10$.
Compared to other studies using scintillation arcs, we note that our observations have a lower $S/N$ ratio.
Most of the power in the secondary spectrum is close to its center as seen in the right panel of Figure \ref{fig:fig1}.

\subsection{Arc-Curvature Likelihoods}\label{chapt:Arc Curvature Likelihoods}

\begin{figure}
	\includegraphics[width=1\columnwidth]{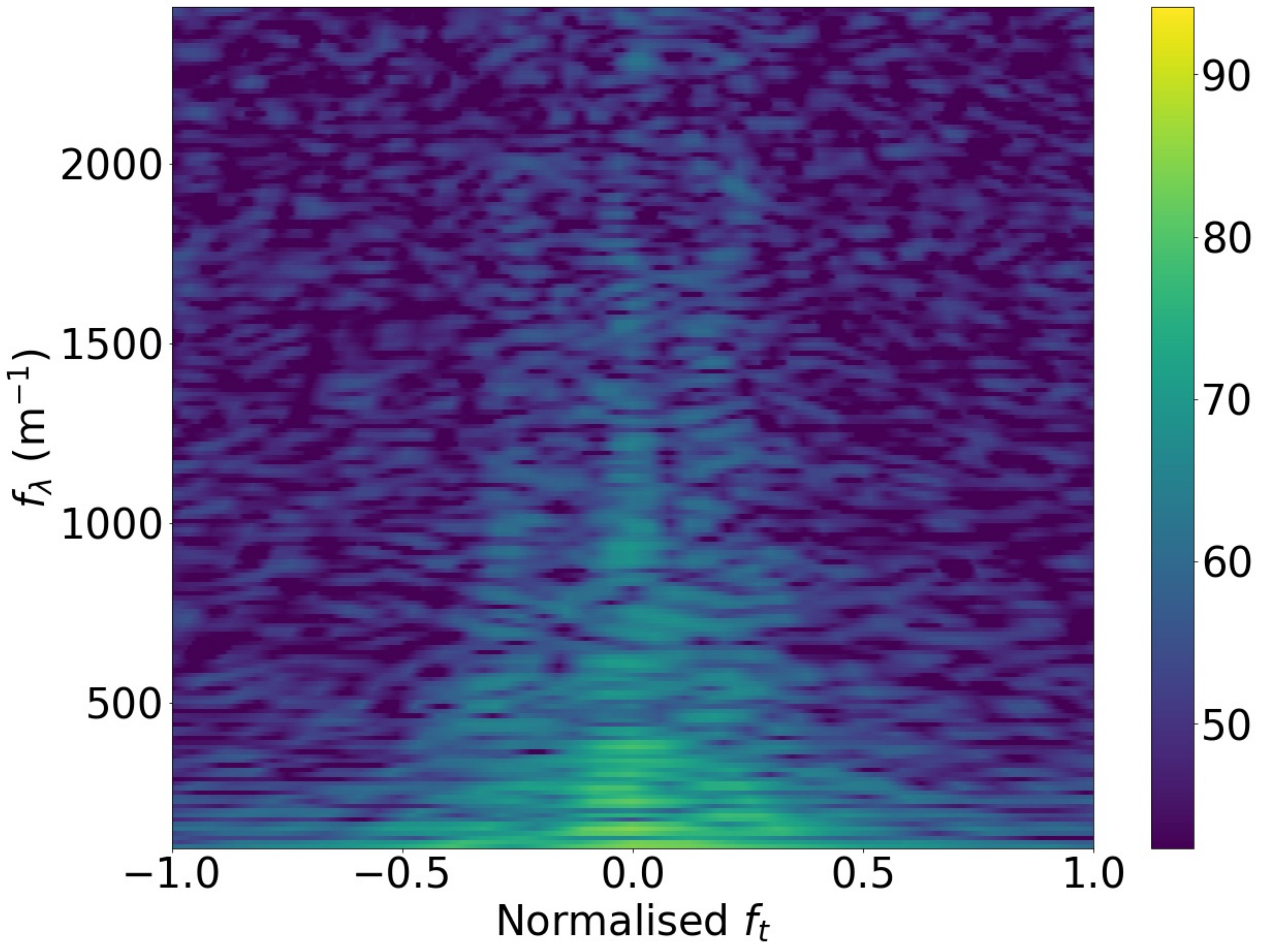}
    \caption{ Normalised secondary spectrum. The power is shown against the Fourier conjugate of $\lambda$ and the normalised Fourier conjugate of time (with respect to reference curvature 100\,mHz$^{-2}$\,m$^{-1}$ in the secondary spectrum. This is the same observation as in Figure \ref{fig:fig1}.}
    \label{fig:normsspec}
\end{figure}

\begin{figure*}
	\includegraphics[trim=50 50 50 50,width=1\columnwidth]{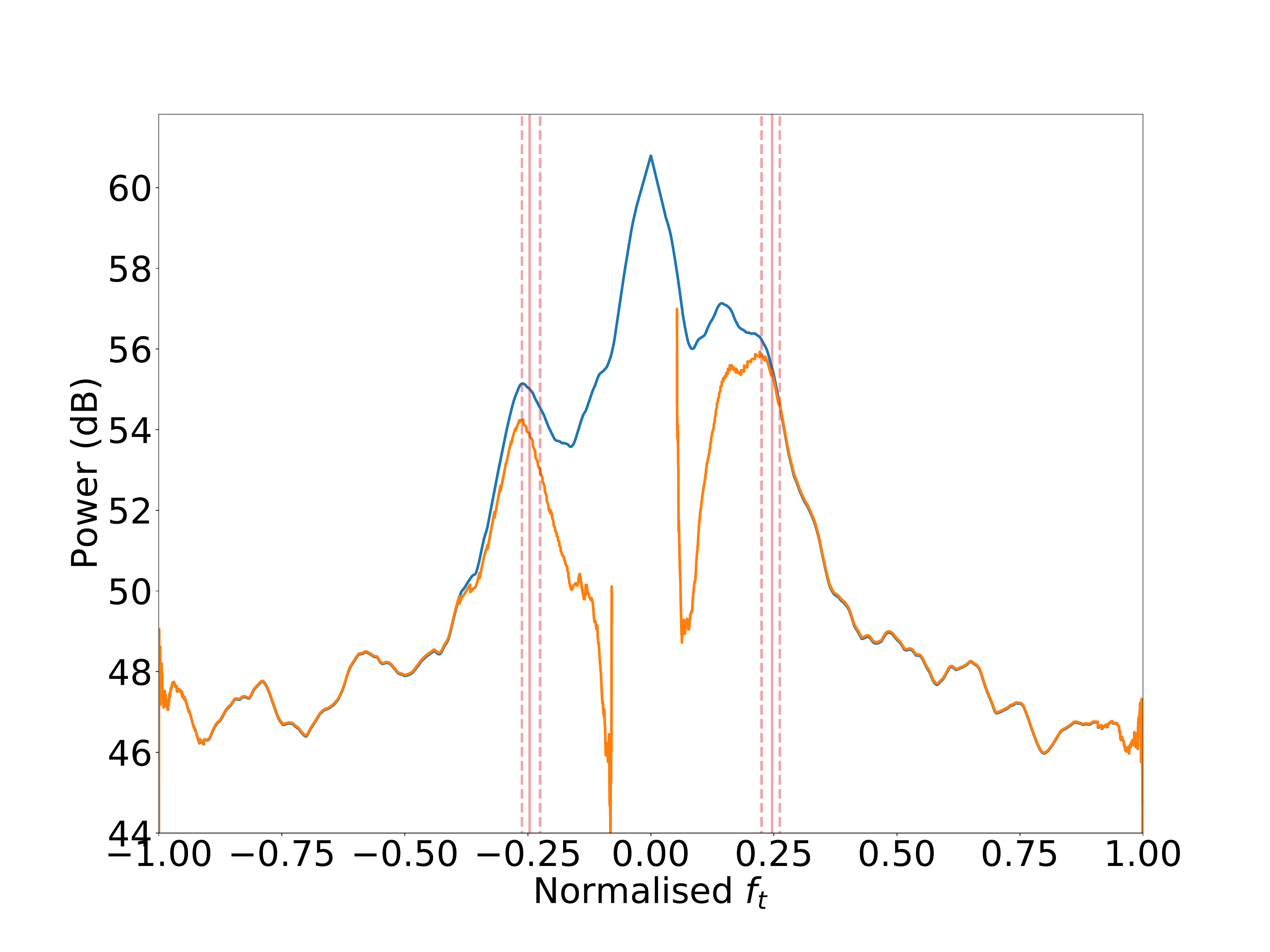}
	\includegraphics[trim=50 50 50 50,width=1\columnwidth]{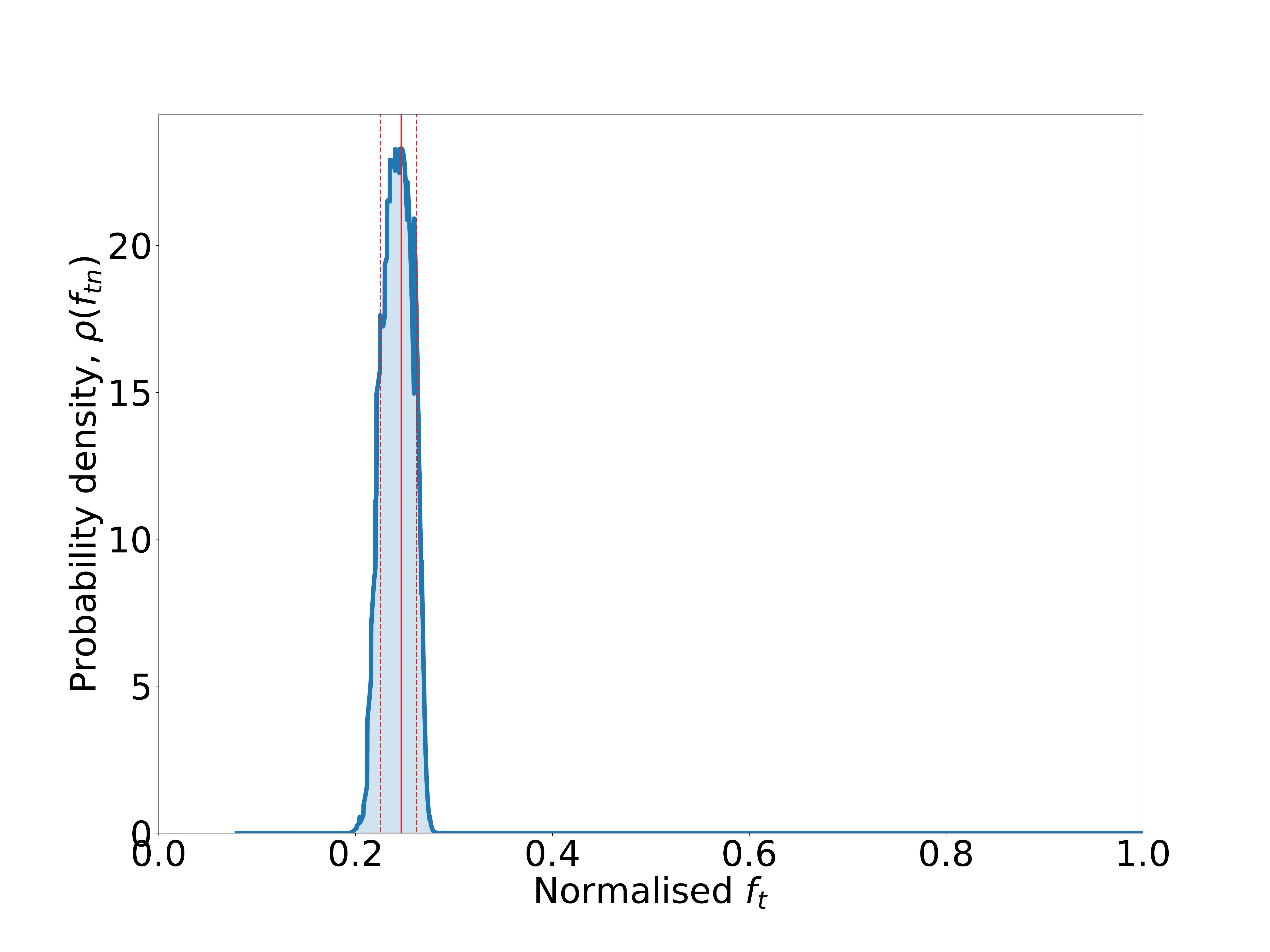}
    \caption{ Arc-curvature determination. This is the same observation as in Figure \ref{fig:fig1}, with summed power (left), and as a normalised PDF (right). For the Doppler profile (left) the blue line indicates the unfiltered power in the secondary spectrum (y-axis) at different values of $f_{tn}$ (x-axis). Whereas, the orange line shows the filtered secondary spectrum after removing noise from the central three pixels along the $f_\lambda$ axis. The red vertical lines indicated the maximum likelihood measurement of $f_{tn}$. The offset of the red line fit with the peaks in the orange line is due to known asymmetries about the $f_{tn}$ = 0 axis, as seen in Section \ref{chapt:Phase Gradients}.}
    \label{fig:DopplerandPDF}
\end{figure*}

Following \citet{Reardon2020} we calculate the normalised secondary spectrum, $P(f_t/f_{\textrm{arc}},f_\lambda)$ (Figure \ref{fig:normsspec}).
This is accomplished by re-sampling the secondary spectrum such that parabolas are straight lines along constant values for normalised conjugate time, $f_t/f_{\textrm{arc}}=f_{tn}$. 
We choose $f_{\textrm{arc}} = 100$\,mHz$^{-2}$\,m$^{-1}$ and then consider only $|f_{tn}|<1$ as the data, as the observed arc curvatures are always greater than this $f_{\textrm{arc}}$.
The Doppler profile (Figure \ref{fig:DopplerandPDF}) $D_t(f_{tn})$ is formed by performing a weighted sum along $f_\lambda$, and shows the power as a function of $f_{tn}$.
One method to measure $\eta$ is to use the peak in maximum power in the Doppler profile ($D_{t,\textrm{max}}$), which has been done for previous studies \citep{Main2020, Reardon2020}.
This could lead to inaccuracies in our measurements because we often see multiple peaks in the Doppler profile, which could be independent arcs or due to noise artefacts.

We estimate the probability of an accurate measurement of $\eta$ by transforming the Doppler profile into a probability density function (PDF) \citep{Walker2022}, as shown in Figure \ref{fig:DopplerandPDF}. 
We first average the positive and negative $f_{tn}$ curves, and calculate the probability density of each $f_{tn}$ being the peak in power assuming a Gaussian likelihood.
This can be expressed as
\begin{eqnarray}
    \rho(f_{tn}) = \frac{1}{\sigma_P\sqrt{2\pi}}e^{-0.5((D_t(f_{tn}) - D_{t,\textrm{max}})/\sigma_P)^2},
    \label{eqn:norm_pdf}
\end{eqnarray}
where
\begin{eqnarray}
    \sigma_P = \sqrt{(\sigma_s \times 10^F)^2 + (10^Q)^2},
    \label{eqn:sigma}
\end{eqnarray}
$\sigma_s$ is the noise in the secondary spectrum, and $\sigma_P$ is the new estimated noise level which is computed using two white noise parameters, $F$ and $Q$, which are included in our model \citep{Walker2022}.

\subsection{Modelling}\label{chapt:Modelling}

We use a Bayesian approach to our modelling while following methods published in \citet{Reardon2020} and \citet{Walker2022}.
For our modelling, we utilise Bayesian inference software \emph{BILBY} \citep{Ashton2019}.
We take our data in the form of $D_t(f_{tn})$ for each observation and transform these into PDFs given a $\sigma_P$.
The product of these PDFs at a model $f_{tn}$ is used to determine the model likelihood \citep{Walker2022}.
We use the \emph{dynesty} sampler to sample the posterior distribution for all the models \citep{Speagle2020}.
This also allows us to calculate the Bayesian ``evidence", $Z$, to select a preferred model for our arc-curvature measurements.
To compare models we use the ratio of evidences called the Bayes factor, ${\rm BF}=\frac{Z_A}{Z_B}$, where $Z_A$ and $Z_B$ are the evidences from models $A$ and $B$, respectively.
In this work, we use the log of the ${\rm BF}$ when comparing evidence values $\log{\rm BF}=\log(Z_A) - \log(Z_B)$, where we define one model to have ``strong evidence'' over the other when $|\,\log{\rm BF}\,|>8$ \citep{Thrane2019}.
In the case where test model $A$ is preferred over base model $B$ the $\log{\rm BF}$ will be positive.

We consider models for arc-curvature measurements that depend on the structure of the IISM and its relative motion to the pulsar and Earth.
The effective velocity is
\begin{eqnarray}
    \textbf{V}_{\textrm{eff}}(s) &&= (1-s)(\textbf{V}_p + \textbf{V}_\mu) + s\textbf{V}_{\textrm{E}} - \textbf{V}_{\textrm{IISM}} \nonumber \\
     &&= \textbf{V}_{\rm kin} - \textbf{V}_{\textrm{IISM}},
	\label{eq:veff}
\end{eqnarray}
where $\textbf{V}_p$ is the pulsars orbital velocity, $\textbf{V}_\mu$ is the pulsar transverse space velocity, $\textbf{V}_{\textrm{E}}$ is Earth's velocity, and $\textbf{V}_{\textrm{IISM}}$ is the IISM velocity.
While the pulsar's proper motion and radial component of orbital velocity are well determined from pulsar timing, the IISM velocity can only be measured through scintillation.
The distance to the pulsar 1152\,pc and the magnitude of proper motion of the pulsar 37.02\,mas\,yr$^{-1}$ were used as fixed values in the modeling.
From our analysis, a number of parameters can be estimated.
This includes the longitude of ascending node, $\Omega$ and inclination angle, $i$ (including the sense of the pulsars orbit).

We also estimate the relative distance to the scattering screen, $s$, which impacts the relative contribution of the Earth and the pulsar to arc-curvature variations.
For isotropic scattering screens, we decompose the IISM velocity into components in direction of constant right ascension ($V_{\textrm{IISM},\alpha}$) and declination ($V_{\textrm{IISM},\delta}$).
In this case, the total effective velocity is
\begin{eqnarray}
    V_{\textrm{eff}}(s) = \sqrt{ \left( V_{\textrm{kin},\alpha} - V_{\textrm{IISM},\alpha} \right)^2 + \left( V_{\textrm{kin},\delta} - V_{\textrm{IISM},\delta} \right)^2}.
	\label{eq:isot_veff}
\end{eqnarray}
For anisotropic models, we parameterize the screens using an anisotropy angle ($\zeta$) and the velocity in this direction ($\textbf{V}_{\textrm{IISM},\zeta}$). 
Scintillation arc curvatures are insensitive to motion perpendicular to the anisotropy.
In this other case, the total effective velocity is
\begin{eqnarray}
    V_{\textrm{eff}}(s) \cos(\psi) = \sqrt{{\left(V_{\textrm{kin},\alpha}\sin{\zeta} + V_{\textrm{kin},\delta}\cos{\zeta}-V_{\textrm{IISM}_{\zeta}}\right)}^2}.
	\label{eq:anis_veff}
\end{eqnarray}

\begin{figure*}
	\includegraphics[trim=50 50 50 50,width=1\columnwidth]{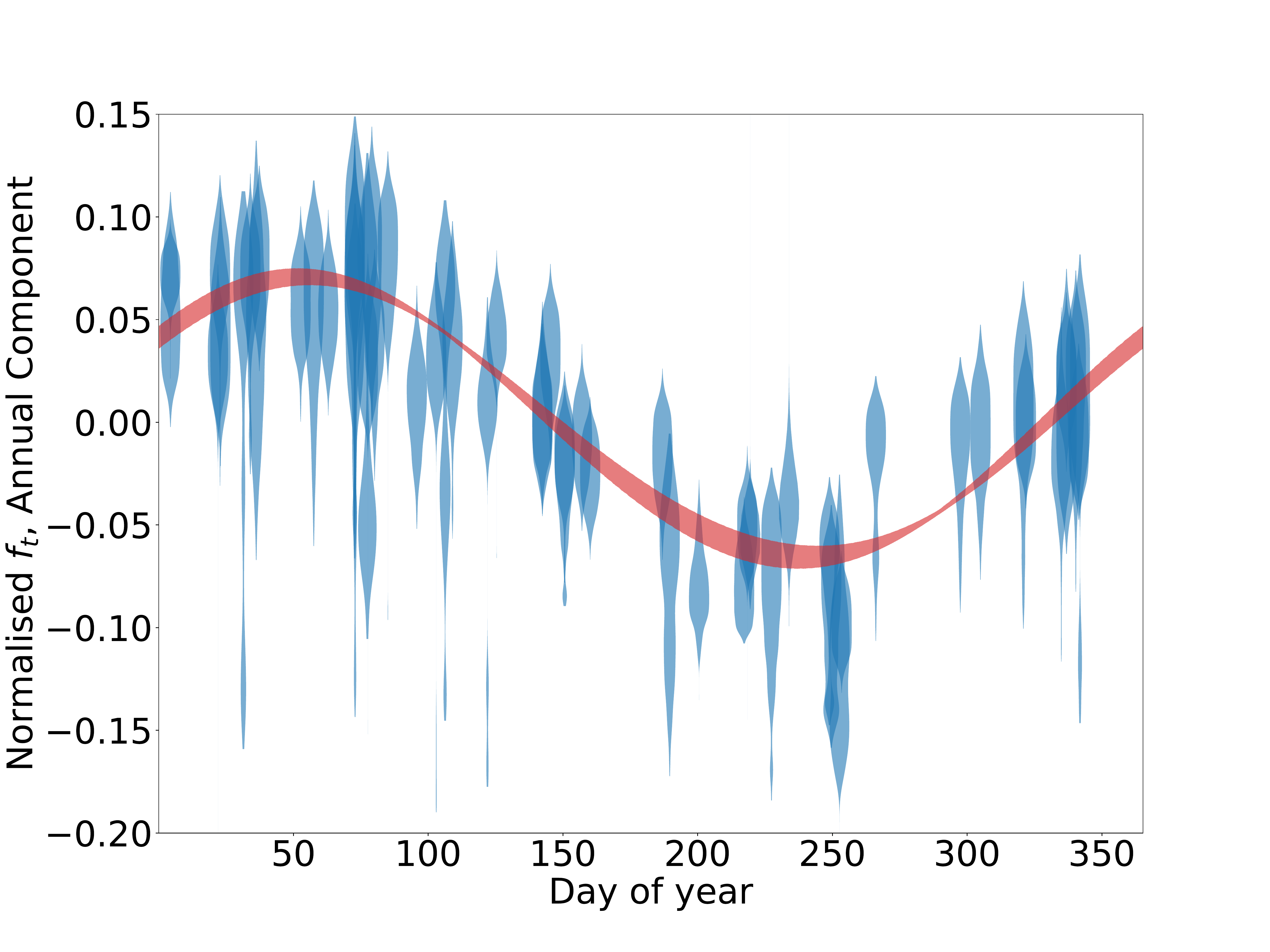} 
	\includegraphics[trim=50 50 50 50,width=1\columnwidth]{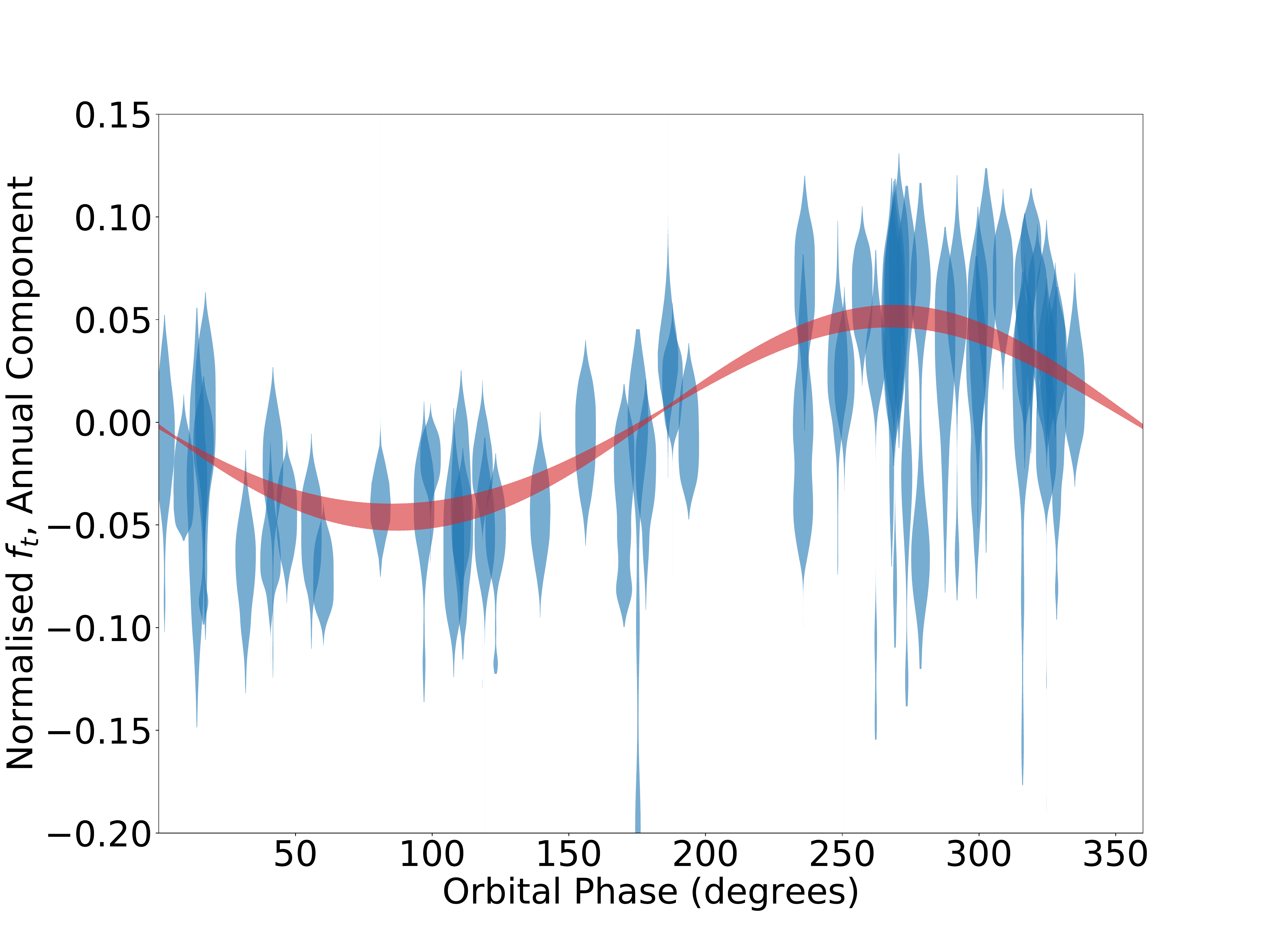} \\
	\includegraphics[trim=50 50 50 50,width=2\columnwidth]{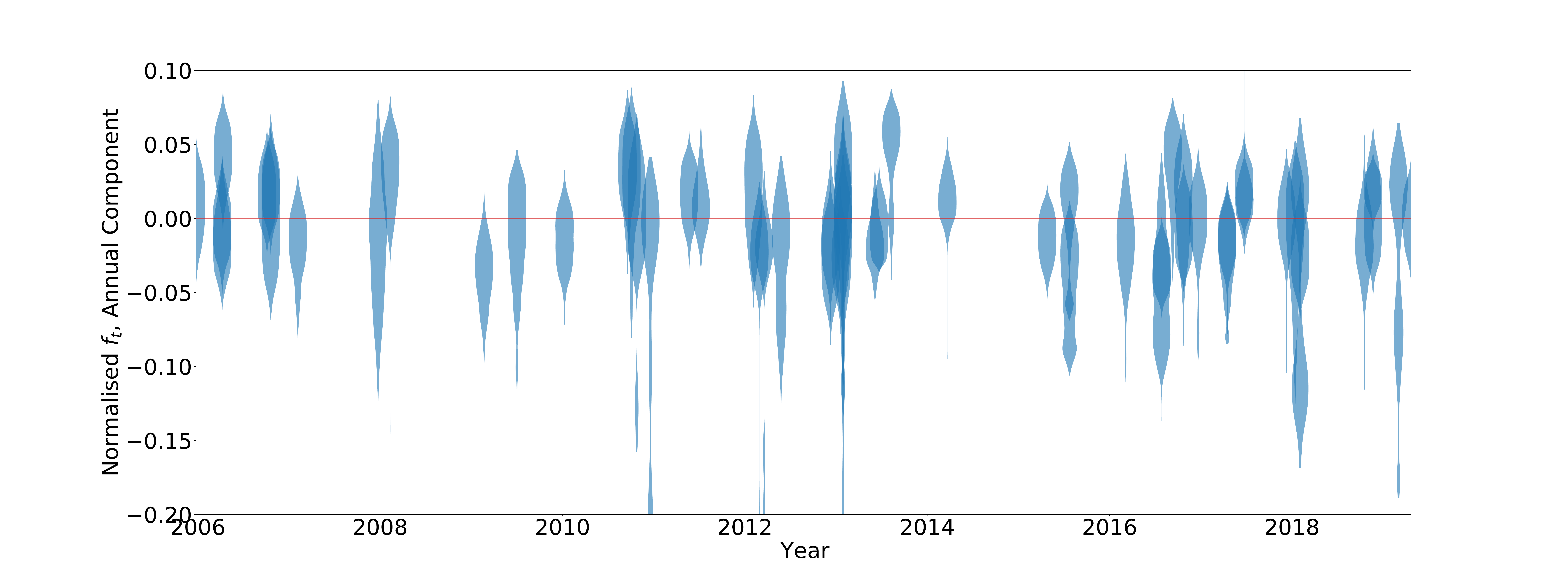} \\
	\includegraphics[trim=50 50 50 50,width=1\columnwidth]{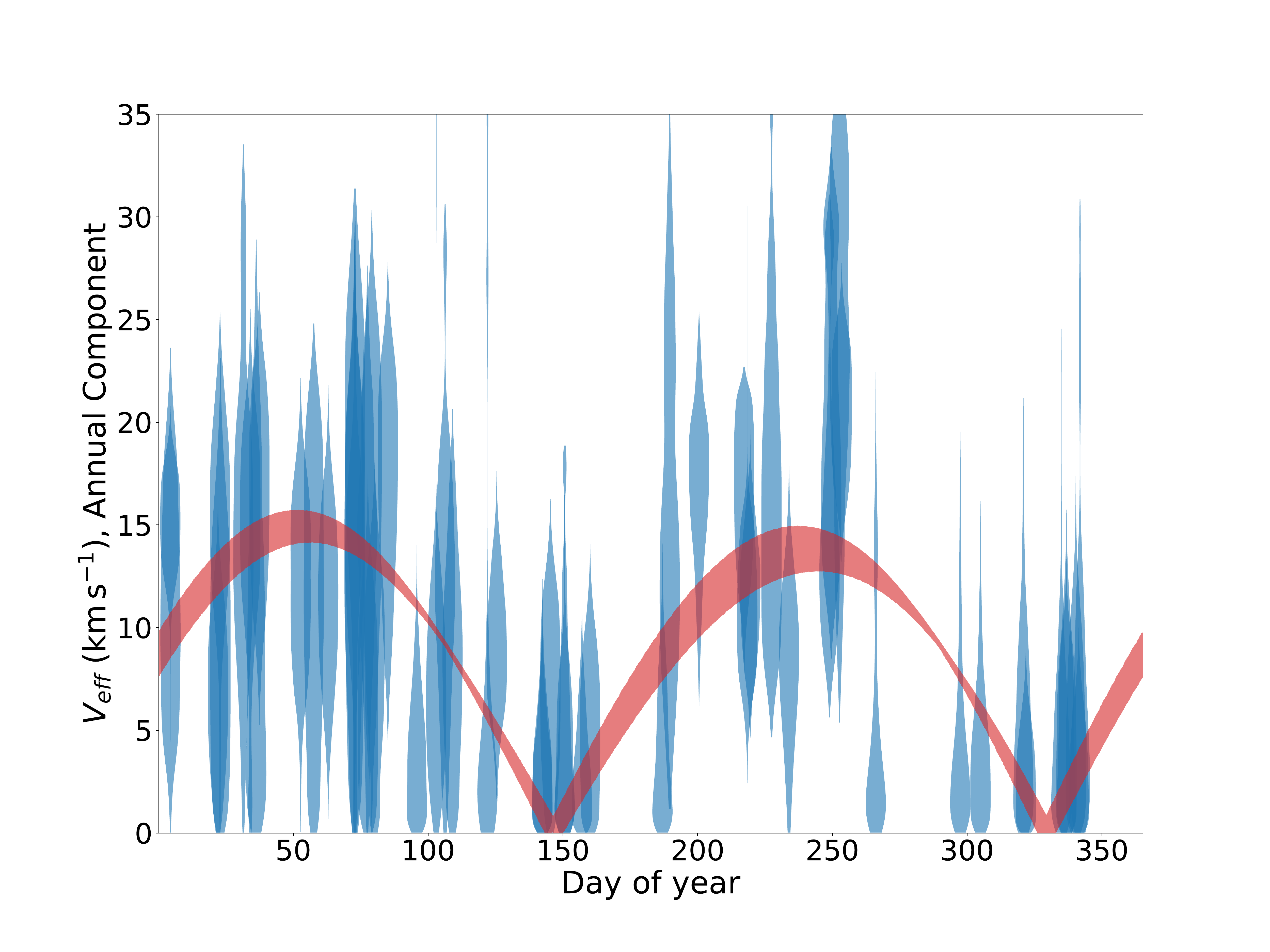} 
	\includegraphics[trim=50 50 50 50,width=1\columnwidth]{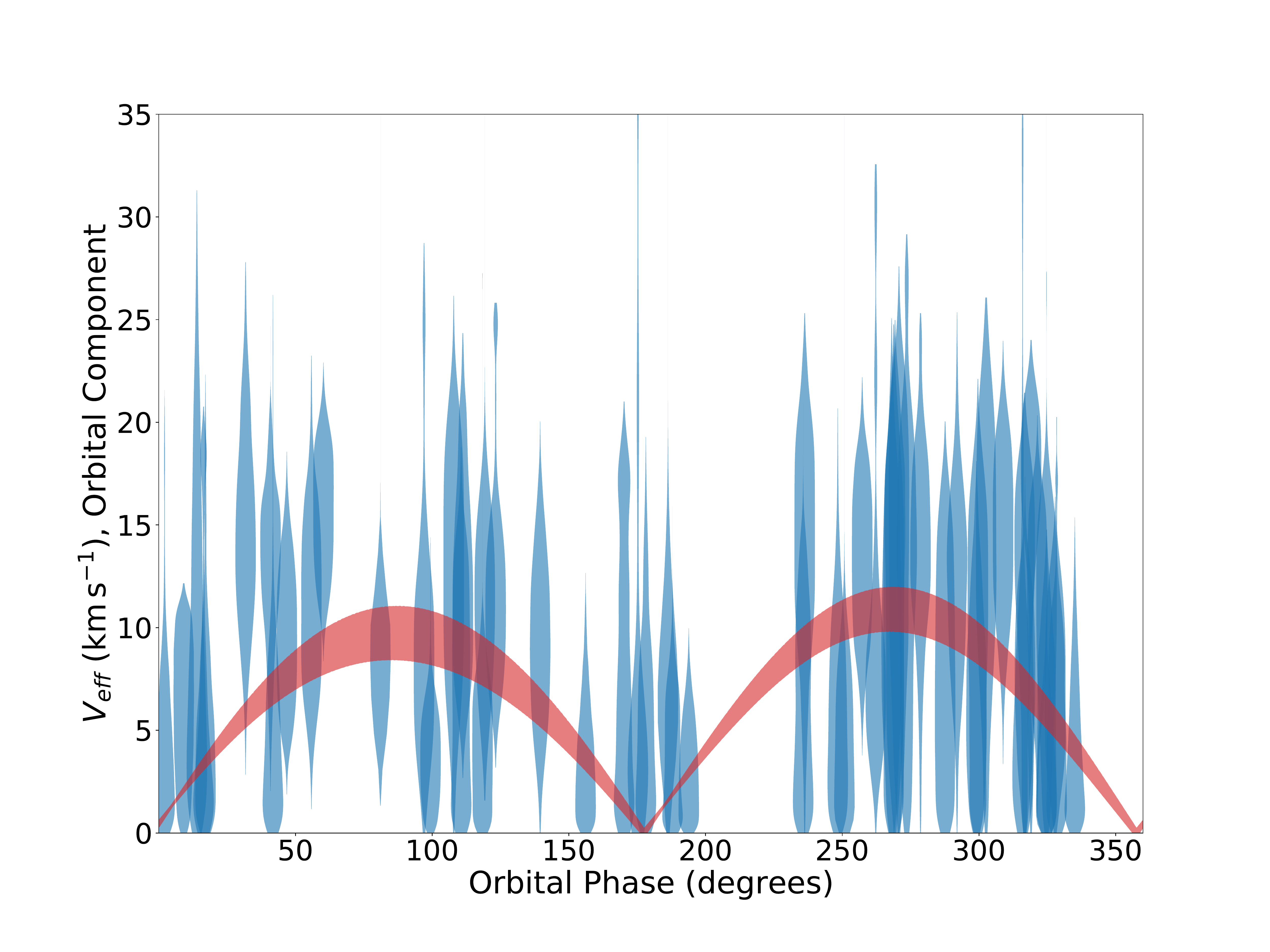} 
    \caption{ Residual arc-curvature measurements for the isotropic model. The top panels show the model subtracted violin plots in blue of the data, with the isotropic model in red. We show the annual variations in the top left panel, the orbital variations in the top right, and the residuals against MJD in the bottom panel. The top left panel shows the annual modulation after subtracting the orbital component from the model, and the top right panel shows the orbital modulation with the annual component subtracted. The bottom two panels show $\boldsymbol{V}_{\textrm{eff}}$ calculated from the annual and orbital modulation models (Equation \ref{eq:veff}). The 2D posterior for this model is shown in Figure \ref{fig:corner_isot}.}
    \label{fig:model_isot}
\end{figure*}

We calculate PDFs as described above in Equation \ref{eqn:norm_pdf}.
We include the two white noise parameters $F$ and $Q$ as defined in Equation \ref{eqn:sigma}.
These parameters were given a uniform prior between $-2$ and $+2$. 
The PDFs of arc curvature for each observation are shown with a ``violin plot'' e.g. Figure \ref{fig:model_isot} (see also \citet{Walker2022}).
Violin plots are PDFs plotted horizontally.
The width of the violin is proportional to the probability density.
We inspect the quality of our models using the residuals.
A good model for the data will show flat residuals, a probability-weighted mean of zero, and each violin intersecting zero (because they represent the full PDF of each observation).

\begin{figure*}
	\includegraphics[trim=50 50 50 50,width=1\columnwidth]{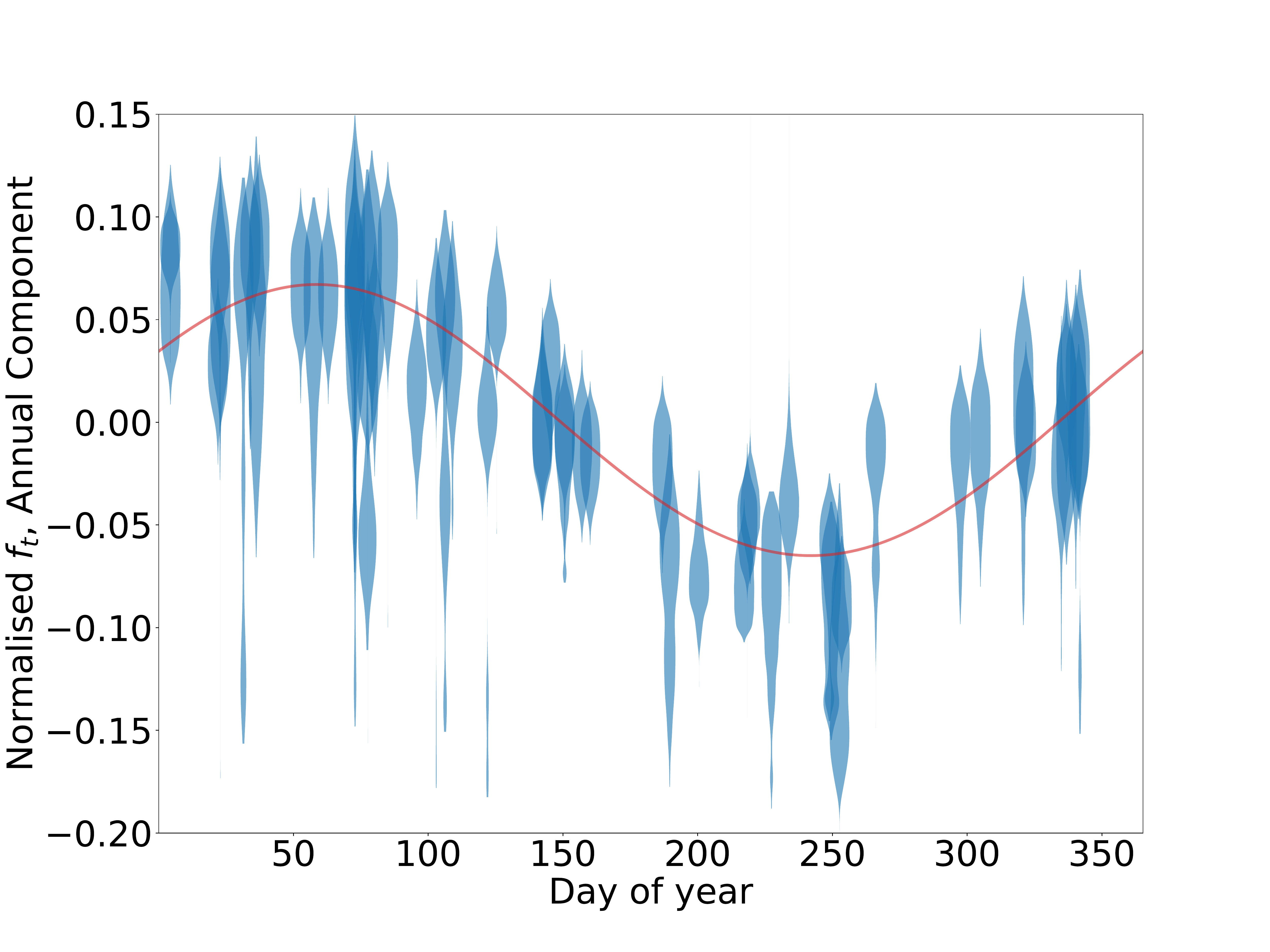} 
	\includegraphics[trim=50 50 50 50,width=1\columnwidth]{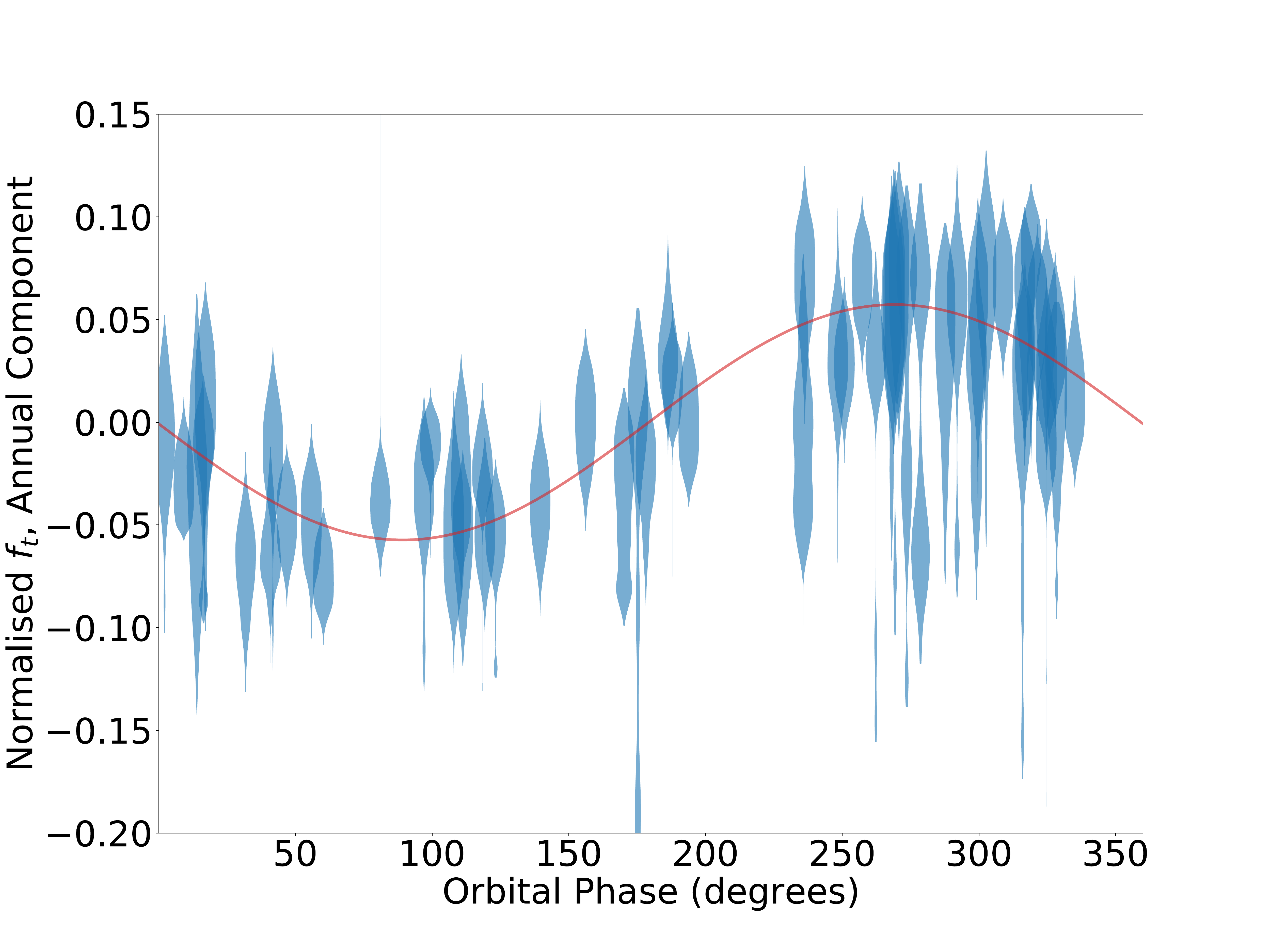} \\
	\includegraphics[trim=50 50 50 50,width=2\columnwidth]{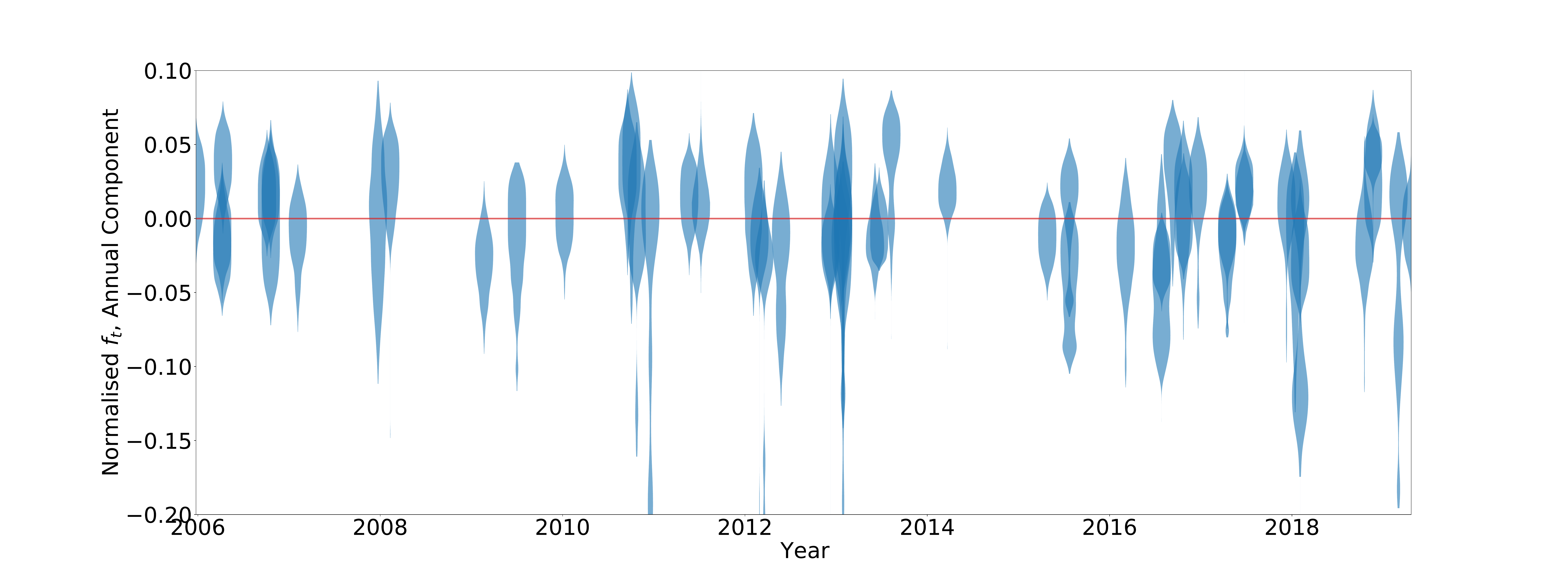} \\
	\includegraphics[trim=50 50 50 50,width=1\columnwidth]{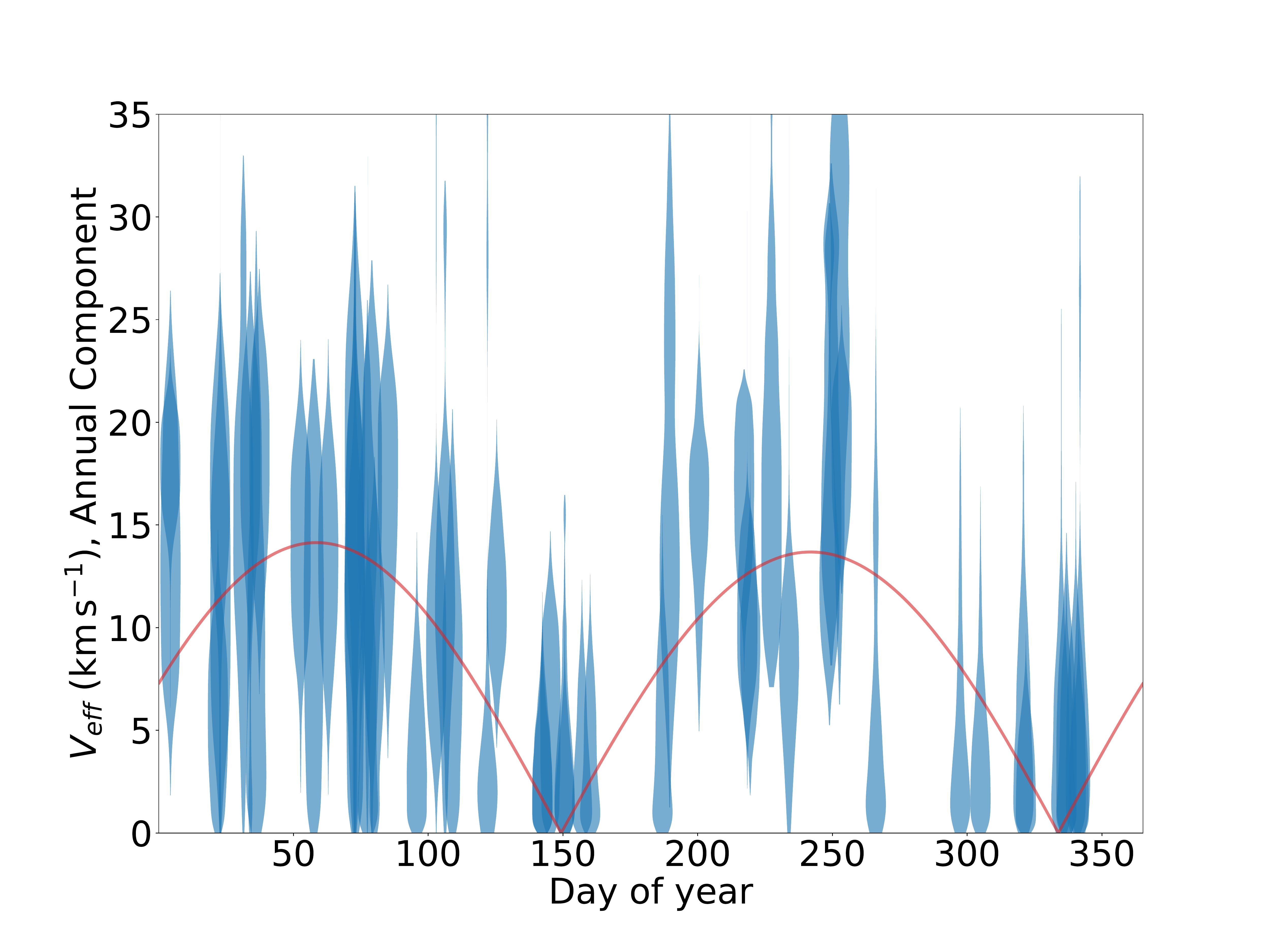} 
	\includegraphics[trim=50 50 50 50,width=1\columnwidth]{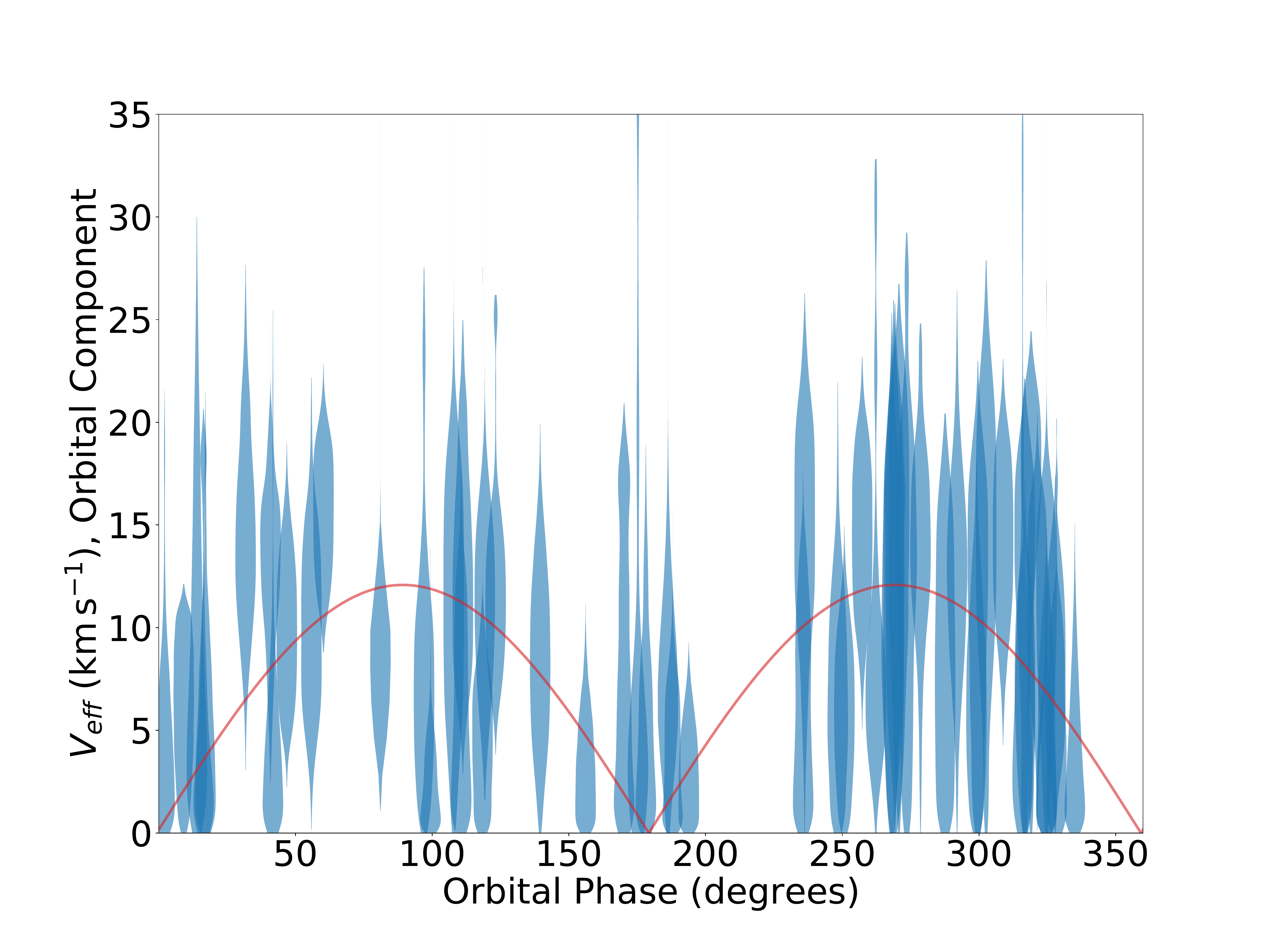} 
    \caption{ Residual arc-curvature measurements for the anisotropic model, following Figure \ref{fig:model_isot}. The 2D posterior for this model is shown in Figure \ref{fig:corner_anis}}
    \label{fig:model_anis}
\end{figure*}

\section{Results}\label{chapt:Results}

\subsection{Scintillation Arc Modelling}\label{chapt:Scintillation Arc Modelling}

We first considered models where we allowed $\Omega$ to vary freely and $i$ to take values consistent with one of the two ambiguous solutions from pulsar timing.
The high precision measurement of $\sin{i}=0.99809\pm0.00004$, from timing, allows for two possible solutions $i=86.46\pm0.05^\circ$ and $i=93.54\pm0.05^\circ$.
The circular orbit of PSR~J1909$-$3744 results in four degenerate solutions for $\Omega$, two for each sense of inclination angle \citep{Reardon2021}.
This is accounted for by creating models for both possible values of $i$.
Our models are sensitive to the annual and orbital variations of $\eta$, which can in principle allow us to constrain $\Omega$.

We then considered isotropic and anisotropic models using the method explained in Section \ref{chapt:Modelling} and shown in Figures \ref{fig:model_isot} and \ref{fig:model_anis}. 
All of these cases were consistent with a scattering screen approximately halfway between Earth and the pulsar at a distance $D_s\approx530\pm50$\,pc. 
The models predict IISM velocities ranging from approximately 20 to 110\,km\,s$^{-1}$, which we discuss further below. 
We find that the inferred values of $\Omega$ are at minimum $2.5\sigma$ from that determined from pulsar timing \citep{Liu2020, Reardon2021}. 
However, in two cases our posterior distributions for $\Omega$ have tails that are skewed away from pulsar timing inferred values. 
All models gave similar log-evidence values.
As such, based on Bayesian evidence alone we cannot identify a preferred model.
The inferred parameters from the models can be found in Table \ref{tab:ourmodel_table}. 
Two-dimensional posterior distributions for isotropic and anisotropic solutions assuming $i=86.46^\circ$ are shown in Figures \ref{fig:corner_isot} and \ref{fig:corner_anis}. 
\begin{table}
	\centering
	\caption{ Scintillation properties from our curvature measurements. The full 2D parameter distributions are shown in Figures \ref{fig:corner_isot} and \ref{fig:corner_anis}.}
	\label{tab:ourmodel_table}
	\begin{tabular}{l | c c c r}
		\hline
	    \multicolumn{1}{|l|}{Fitted} &
	    \multicolumn{2}{|c|}{$i$ = 86.46$^\circ$} &
	    \multicolumn{2}{|c|}{$i$ = 93.54$^\circ$} \cr
	    Parameter & Isotropic & Anisotropic & Isotropic & Anisotropic \\
        \hline
        $s$ & $0.53_{-0.04}^{+0.04}$ & $0.54_{-0.04}^{+0.04}$ & $0.55_{-0.04}^{+0.04}$ & $0.54_{-0.04}^{+0.04}$ \\
        \\
        $\Omega$ ($^\circ$) & $320_{-10}^{+8}$ & $310_{-30}^{+10}$ & $310_{-40}^{+20}$ & $270_{-30}^{+40}$ \\
        \\
        $V_{\textrm{IISM},\alpha}$ (km\,s$^{-1}$) & $21_{-2}^{+2}$ & $-$ & $20_{-2}^{+2}$ & $-$ \\
        \\
        $V_{\textrm{IISM},\delta}$ (km\,s$^{-1}$) & $-107_{-6}^{+6}$ & $-$ & $106_{-5}^{+6}$ & $-$ \\
        \\
        $\zeta$ ($^\circ$) & $-$ & $113_{-10}^{+7}$ & $-$ & $112_{-10}^{+8}$ \\
        \\
        $\textbf{V}_{\textrm{IISM}_{\zeta}}$ (km\,s$^{-1}$) & $-$ & $60_{-20}^{+10}$ & $-$ & $60_{-20}^{+10}$ \\
        \\
		\hline
        log evidence & -367.6(1) & -367.5(1) & -368.6(1) & -367.6(1) \\
        \hline
	\end{tabular}
\end{table}

\begin{table*}
	\centering
	\caption{ Scintillation properties inferred using timing measurements of $\Omega$. We have assumed an inclination angle of $i=86.46^\circ$.}
	\label{tab:literaturemodel_table}
	\begin{tabular}{l | c c c c c c c r}
		\hline
	    &  \multicolumn{4}{|c|}{\citet{Reardon2021}} &  \multicolumn{4}{|c|}{\citet{Liu2020}} \\
	    \multicolumn{1}{|l|}{$\pi(\Omega)$} &
	    \multicolumn{2}{|c|}{$\Omega=225\,\pm\,3^\circ$} &
	    \multicolumn{2}{|c|}{$\Omega=344\,\pm\,3^\circ$} &
	    \multicolumn{2}{|c|}{$\Omega=217\,\pm\,5^\circ$} &
	    \multicolumn{2}{|c|}{$\Omega=352\,\pm\,5^\circ$} \\
	    Parameter & Isotropic & Anisotropic & Isotropic & Anisotropic & Isotropic & Anisotropic & Isotropic & Anisotropic \\
        \hline
        $s$ & $0.49_{-0.04}^{+0.03}$ & $0.49_{-0.03}^{+0.04}$ & $0.54_{-0.04}^{+0.04}$ & $0.54_{-0.04}^{+0.04}$ & $0.49_{-0.04}^{+0.04}$ & $0.49_{-0.03}^{+0.03}$ & $0.54_{-0.04}^{+0.04}$ & $0.54_{-0.04}^{+0.04}$ \\
        \\
        $\Omega_{\rm post}$  ($^\circ$) & $225_{-3}^{+3}$ & $226_{-3}^{+4}$ & $340_{-3}^{+3}$ & $340_{-3}^{+3}$ & $218_{-5}^{+5}$ & $220_{-5}^{+5}$ & $340_{-5}^{+4}$ & $340_{-4}^{+4}$ \\
        \\
        $V_{\textrm{IISM},\alpha}$ (km\,s$^{-1}$) & $21_{-2}^{+2}$ & $-$ & $18_{-2}^{+2}$ & $-$ & $20_{-2}^{+2}$ & $-$ & $18_{-3}^{+2}$ & $-$ \\
        \\
        $V_{\textrm{IISM},\delta}$ (km\,s$^{-1}$) & $-90_{-9}^{+9}$ & $-$ & $-112_{-7}^{+6}$ & $-$ & $-85_{-9}^{+10}$ & $-$ & $-112_{-7}^{+6}$ & $-$ \\
        \\
        $\zeta$ ($^\circ$) & $-$ & $85_{-6}^{+5}$ & $-$ & $120_{-5}^{+5}$ & $-$ & $81_{-8}^{+6}$ & $-$ & $120_{-5}^{+5}$ \\
        \\
        $\textbf{V}_{\textrm{IISM}_{\zeta}}$ (km\,s$^{-1}$) & $-$ & $14_{-10}^{+8}$ & $-$ & $72_{-7}^{+6}$ & $-$ & $6_{-10}^{+9}$ & $-$ & $72_{-7}^{+6}$ \\
		\hline
        log evidence & -371.0(1) & -369.5(1) & -370.0(1) & -371.6(1) & -371.3(1) & -369.9(1) & -372.1(1) & -373.0(1) \\
        \hline
	\end{tabular}
\end{table*}

\subsection{Alternative Models}\label{chapt:Alternative Models}

The inferred values from initial scintillation modelling revealed inconsistencies with pulsar timing.
In particular our measurements of $\Omega$ being 2.5$\sigma$ from pulsar timing results of \citet{Liu2020, Reardon2021}.
We also determined unexpectedly high values for $V_{\textrm{IISM},\delta}$ ($>>$10\,km\,s$^{-1}$ see Section \ref{chapt:Velocities of the IISM}).
This indicated something could be incorrect with our assumptions.
Here we investigate possible alternate models that may explain these inconsistencies. 

It is possible that our assumptions that the IISM remains statistically identical over the entire observation is incorrect.
Alternative models of the IISM were explored in this work to investigate the potential for multiple scattering screens along the line of sight.
Scintillation models can provide highly precise measurements of $s$.
Recent works have found that the dominant scattering screen can change, or multiple scattering screens can be present \citep{Reardon2020, Walker2022, Sprenger2022}.
To test this, the data were split into groups divided at an MJD, which was a free parameter.
In each group, the IISM was modelled independently.
We also tested the scenario where $V_{\textrm{IISM}}$ and the properties of the anisotropy could change but location of the scattering screen $s$ was fixed.
All of these models revealed log Bayes factors between $-3$ and $-5$, showing support for the original model.
We conclude that there is weak evidence from the arc-curvature measurements against the presence of multiple scattering screens. 
Our stationary-screen model assumption is therefore not likely to be causing the discrepant values of $\Omega$.
This also suggests the properties of the turbulent IISM do not change dramatically for $\approx$13\,years which is discussed further in Section \ref{chapt:Velocities of the IISM}.

We have also explored the possibility of systematic errors in our modelling methods.
Several generations of back-end processors have been used at Murriyang \citep{Manchester2013}, as technology improved.
To determine if any of the back-ends were impacting our inference, we modelled each back-end independently.
The results from the individual back-ends were combined (i.e., summed the log evidence of all back-end models) to compare with the original model.
Compared to using a single base model for the full dataset, splitting the data by back-end system and fitting a separate model to each was strongly disfavoured with $\log {\rm BF}<-16$.
We also tested for outliers in our dataset using a similar method.
This involved splitting the data into halves (a central MJD of 56162, $\log {\rm BF}<-4$), odd and even pairs ($\log {\rm BF}<-11$), and splitting the data randomly ($\log {\rm BF}<-10$).
This demonstrated that no individual data point was significantly impacting the modelling.
These tests indicated that our initial modelling strategy was not fundamentally flawed and new approaches are needed to explain the anomalous results.

\subsection{Inference using pulsar timing priors}\label{chapt:Inference using pulsar timing priors}

We also conducted our analysis using pulsar timing-derived measurements of $\Omega$.
The four possible values of $\Omega$ from \citet{Liu2020} and \citet{Reardon2021} were used as priors. 
The results of this modelling are summarised in Table \ref{tab:literaturemodel_table}. 

To investigate which sense of $i$ was favoured by our data we tested the models with $i=93.54^\circ$ against base models with $i=86.46^\circ$.
We found $\log{\rm BF}<-13$, which means the data strongly supports models with $i=86.46^\circ$.
However, among the models using $i=86.46^\circ$, no significant difference in Bayesian evidence is seen for different possible values of $\Omega$ and anisotropy.
We do find a variation in the IISM parameters, discussed further in Section \ref{chapt:Velocities of the IISM}.
For one of the solutions presented in \citet{Liu2020}, we find a 2-3$\sigma$ discrepancy between the prior and posterior values of $\Omega$.
We comment on the most likely solution from this method of model comparison and the implications of these results in Section \ref{chapt:Discussion}.

\subsection{Arguments for Anisotropy}\label{chapt:Evidence for Anisotropy and Phase Gradients}

\begin{figure*}
	\includegraphics[width=1\columnwidth]{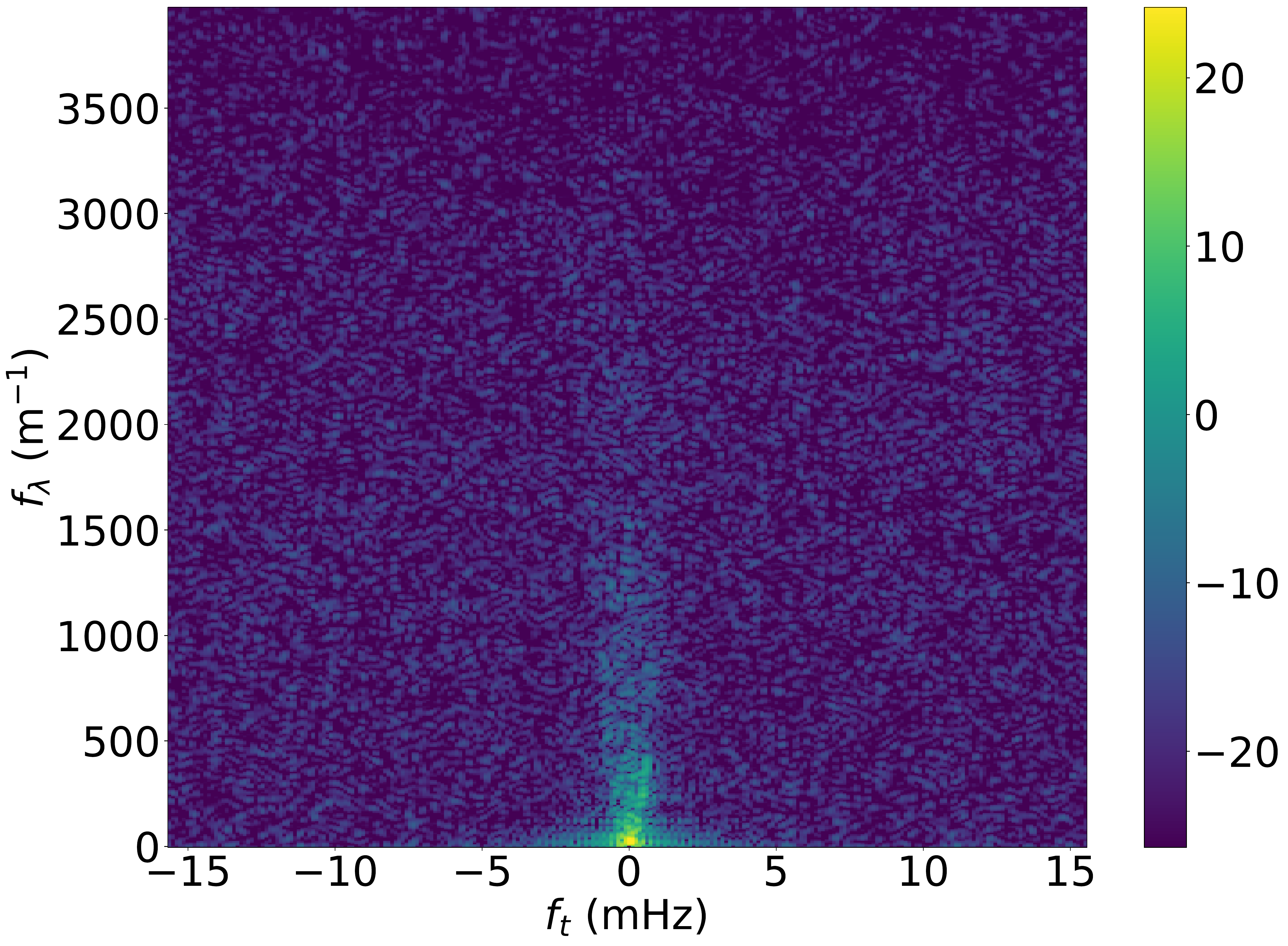}
	\includegraphics[width=1\columnwidth]{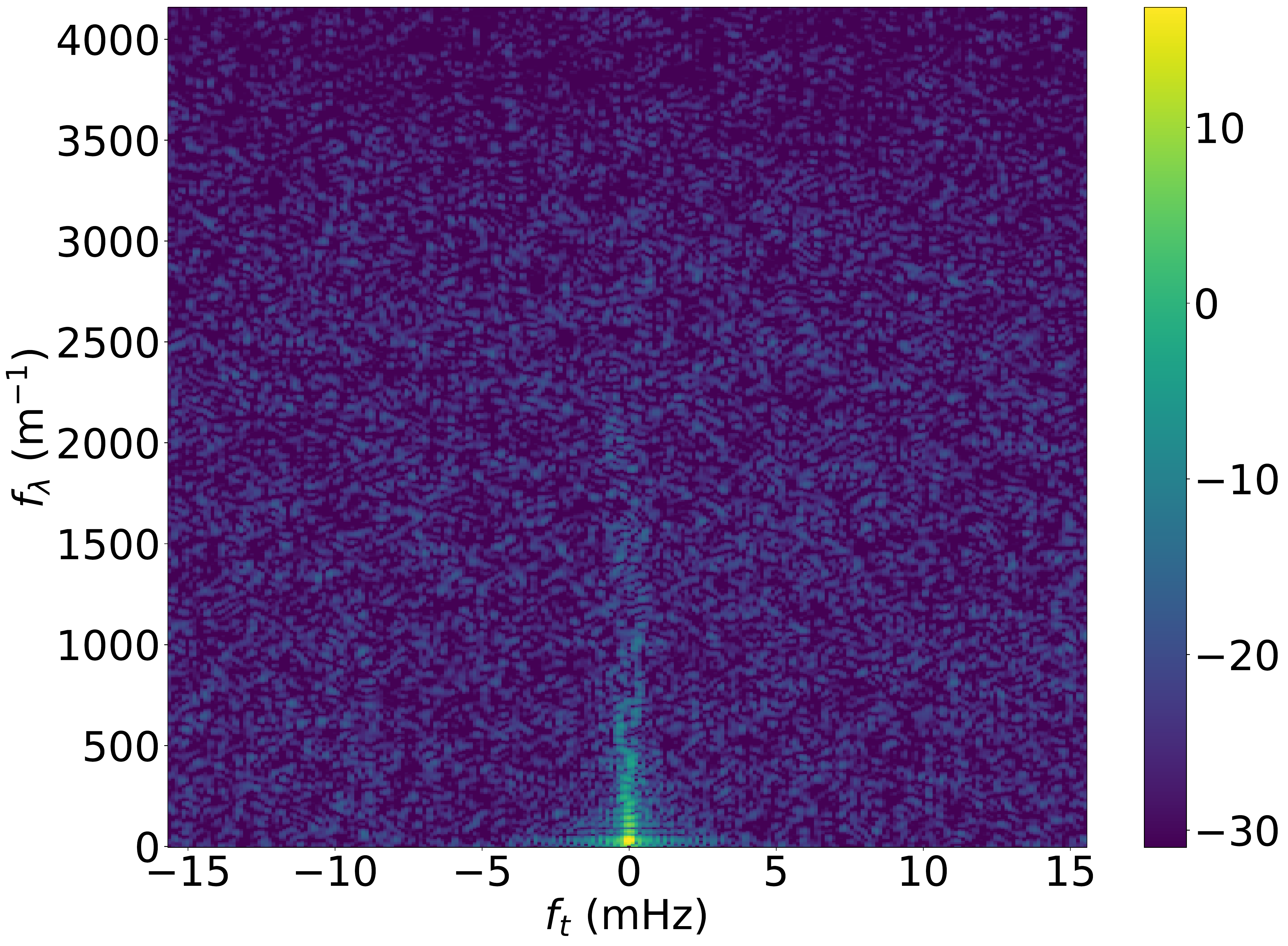} \\
	\includegraphics[width=1\columnwidth]{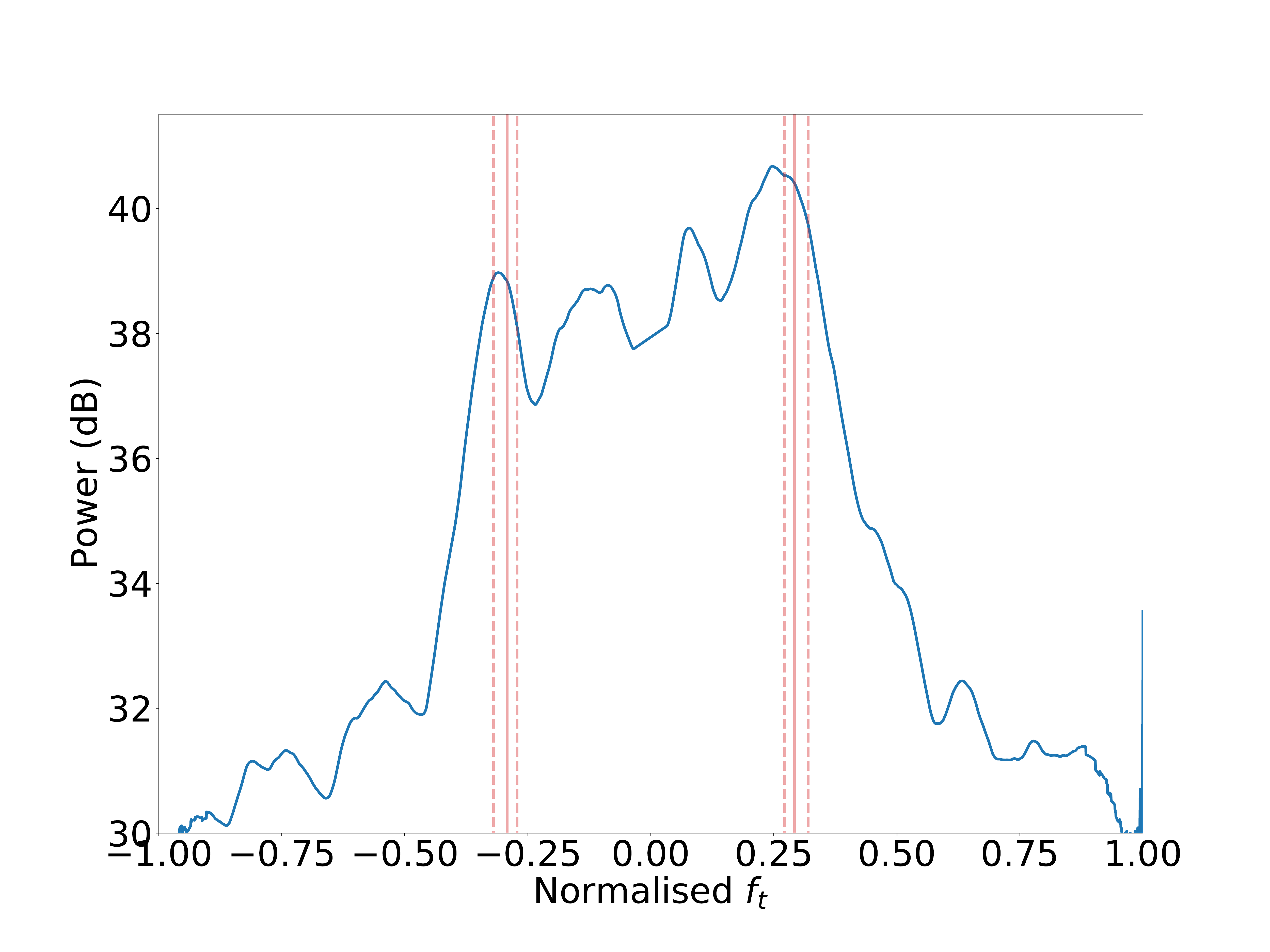}
	\includegraphics[width=1\columnwidth]{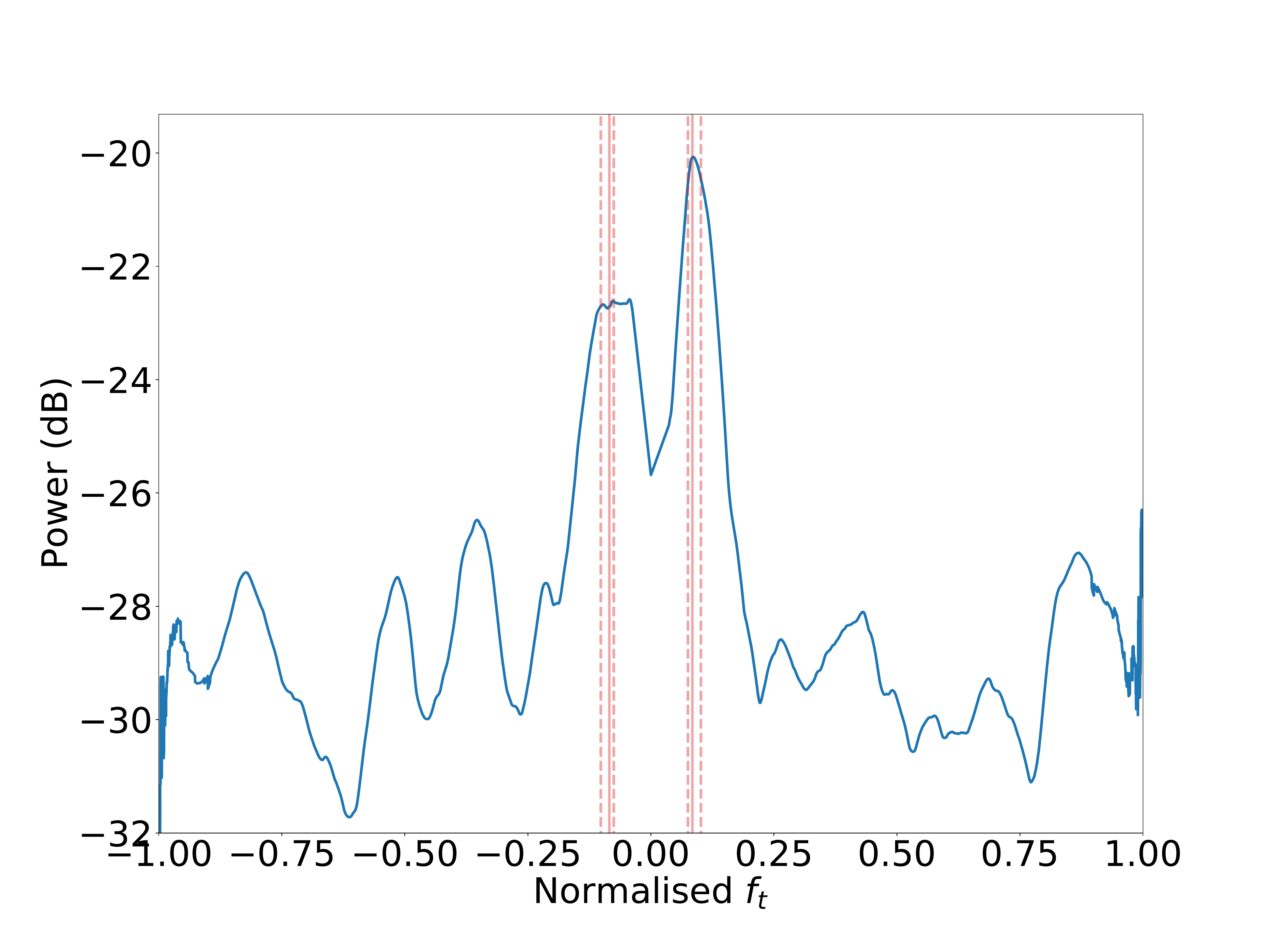}
    \caption{ Arc morphology with different $\psi$. The top panels show the scintillation arcs in the secondary spectrum with observations taken on 2010-12-18 with $\psi=$60.235$^\circ$ (left) and 2016-07-27 with $\psi=$37.670$^\circ$ (right). The difference in arc morphology with $\psi$ is evident in the Doppler profiles (bottom panels). While on the left we observe power distributed interior to the arc-curvature peak, on the right, there is a distinct well of power. These observations were chosen due to their extent in $f_\lambda$ and low level of noise along the $f_t=0$ axis.}
    \label{fig:anis_morphology}
\end{figure*}

While we cannot distinguish between the isotropic and anisotropic scattering using scintillation arc curvatures alone, it is possible to search for anisotropy using arc morphology.
If the arcs arise from anisotropic scattering we would expect to see a change in the arc morphology that depends on $\psi$, the angle between the direction of anisotropy and the effective velocity (Figure \ref{fig:anis_morphology}).
As shown in the appendix of \citet{Reardon2020}, the power in the secondary spectrum is distributed along the arcs in an isotropic model, whereas, as $\psi$ increases from 0$^\circ$ to 90$^\circ$ the power is greater interior to the arc.
For every observation, using our models we calculated that $\psi$ ranges from $38^\circ-68^\circ$.
The observations spanned a sufficient range in $\psi$ to search for evidence of anisotropy in the morphology.
Next, we inspected the power morphology in the Doppler profile (Figure \ref{fig:DopplerandPDF}).
In the case of anisotropic scattering, we would expect there to be two changes to arc morphology.
If $\psi<45^\circ$, there would be a `well' of reduced power interior to the arc-curvature peak.
When $\psi>45^\circ$, then the opposite would be the case so there would be excess power interior to the peak.
A small number ($<$12) of observations were found that satisfied these criteria. 
Anisotropic scattering is also expected to reduce the sharpness of arcs in the secondary spectra.
A majority (90\%) of the observations show broadened distributions of power when they peak in the Doppler profile, expected in anisotropic scattering. 
We cannot definitively see this effect due to the low $S/N$ of our observations. 
Observations with highly anisotropic scattering (axial ratios $A_r\ge2$), aligned with $\boldsymbol{V}_{\textrm{eff}}$, can show scintillation arcs with inverted arclets.
In this highly anisotropic scattering regime, a deep well of power at $f_{tn}=0$ is observed in the arc morphology.
However, we see no evidence for inverted arclets or deep wells of power.
The results here indicate some support for anisotropic models, although the anisotropic scattering is likely to be modest.
We computed using the arc simulation techniques employed by \citet{Reardon2020} \citep[using electromagnetic simulation][]{Coles2010}, that the anisotropic morphology dominates for $A_r\gtrsim1.2$ for this pulsar.
This suggests that the anisotropy is at least at this level, but is unlikely to be highly anisotropic.
Axial ratios of order $\sim$2-3 have been inferred in similar analyses for other pulsars \citep{Rickett2014, Reardon2019}.
The implications of anisotropic scattering are discussed further in Section \ref{chapt:Constraining Anisotropy}.

\subsection{Observation-Observation Parameter Variations}\label{chapt:Observation-Observation Parameter Variations}

Within our modelling, we measured values of $V_{\textrm{IISM}}$ that were inconsistent with expected values, as discussed in Section \ref{chapt:Velocities of the IISM}.
We considered whether short timescale changes in the IISM parameters could be introducing excess noise in our dataset.
Between our observations, the pulsar moves a significant distance.
The pulsar is known to have a transverse velocity $203.2\pm0.5$\,km\,s$^{-1}$, radial velocity $-73\pm30$\,km\,s$^{-1}$ and 3D space velocity of $218\pm10$\,km\,s$^{-1}$, with respect to our solar system barycentre \citep{Liu2020}.
We determine the size of the scattering disk, using $s_r/r_F = \sqrt{\nu_c/\Delta\nu}$ where $r_F=\sqrt{D/k}$ is the Fresnel scale, $k=2\pi/\lambda$ is the wavenumber, and $\Delta\nu\approx14$\,MHz is the median scintillation bandwidth across our measurements \citep{Cordes2001, Reardon2019}.
This gives us $s_r\approx0.05$\,AU implying that each day the pulsar moves outside of the scattering disk ($\approx$0.12\,AU per day).
Throughout our entire observing span, the pulsar has moved $\approx$556.92\,AU.
However, a sufficient amount of time passes between most observations (a mean separation of $\approx$87\,days) such that each observation is interacting with an independent region of the intervening scattering screen.
 
Observation to observation changes in the astrophysical properties of the IISM could lead to changes in $\eta$.
In addition, these would not be fully accounted for in our models described in Section \ref{chapt:Alternative Models}, which consider long-term ($\sim$\,yrs) variations in the IISM.
For example small velocity currents within the turbulent plasma or small changes in anisotropy angle, $\zeta$.
Therefore we can measure the rms of the residuals of $\sigma f_{tn}=0.02$, after subtracting the maximum-likelihood model, and determine how much the IISM parameters would need to vary to match this.
The method and equations used to determine these values are explained further in Appendix \ref{chapt:Observation-Observation Variations in eta methods}.
To account for this the relative distance to the scattering screen would need to vary by $\sigma s=0.06\pm0.01$ (corresponding to a physical size of 70$\pm$10\,pc), in the case of isotropic scattering.
Between observations, the pulsar projected on the screen moves by $2.5\times10^{-5}$\,pc. 
To produce these variations would require a corrugated screen with structures of widths smaller than that but depths of 60\,pc, with the corrugations projected directly towards Earth.
We find this scenario to be unlikely.
In the case of isotropic scattering the velocity variations in right ascension and declination would need to be $\sigma V_{\textrm{IISM},\alpha}=5\pm1\,$km\,s$^{-1}$ and $\sigma V_{\textrm{IISM},\delta}=15\pm3\,$km\,s$^{-1}$, respectively.
In the case of anisotropic scattering, the variation in screen distance is also an implausibly large value of $\sigma s=0.05\pm0.01$ ($58\pm12$\,pc).
The variations in the anisotropic velocity are $\sigma\textbf{V}_{\textrm{IISM}_{\zeta}}=6\pm1\,$km\,s$^{-1}$, with the angle of anisotropic variations being $\sigma\zeta=3\pm1^\circ$.
Our results show the importance of considering these effects when modelling arc-curvature measurements.
Further implications are found within Section \ref{chapt:Velocities of the IISM}.

\subsection{Phase Gradients}\label{chapt:Phase Gradients}

Our data is also sensitive to gradients in electron density.
These are manifested as gradients in phase of the electric field, $\nabla\phi$, which lead to frequency-dependent angular displacements, $\boldsymbol{\theta}_p=\nabla\phi/k$.
These can be observed in many ways, including scintles being askew relative to the frequency-time plane.
In the secondary spectrum, they are manifested as an asymmetry in the power distribution on the positive and negative sides of a scintillation arc \citep{Cordes2006, Rickett1990, Rickett2014}.
Although we do not measure $\nabla\phi$ directly, we see evidence for it within our data.
We can see this in the secondary spectrum if the apex of the parabola has shifted in $f_t$ and $f_\lambda$ (positive or negative, identified in $\approx$10\% of observations).
This effect can be seen clearly in Figure \ref{fig:DopplerandPDF} as the measurement of $f_{tn}$ is not perfectly aligned with the peak in power. 
This effect is also seen when one side of the arc has reduced power (identified in $\approx$40\% of observations).
Due to the prevalence of this effect being found throughout our data set ($\approx$40\%), it was ignored and both sides of $f_{tn}$ were averaged together to determine the most likely measurement of $f_{tn}$.
It is possible that phase gradients could contribute to the excess scatter in $f_{tn}$.
The low sensitivity of the observations makes it difficult to measure the curvature of the arcs independently for positive and negative $f_{tn}$ which could be used to assess excess noise in the arc-curvature measurements.
While our low sensitivity precluded us from measuring $\nabla\phi$, this could be done with more sensitive observations.
The power asymmetry in the arcs could also be used to study in more detail the presence of phase gradients.
These could then, for example, be compared to dispersion measure variations, which can also be used to infer $\nabla\phi$.

\subsection{Time Delay}\label{chapt:Time Delay}

The time delay caused by scattering is often ignored in pulsar timing analysis.
However, it is likely to be a source of minor noise in pulsar timing models \citep{Shannon2017}.
Previous works have used scintillation to measure the scattering time delay $\tau$ including \citet{Hemberger2008, Main2020}.
These methods involved using scintillation arcs to measure the scattering delay.
Instead, we infer the scattering time from the scintillation bandwidth as the $S/N$ of the arcs are low.
To accomplish this we relate $\tau$ to the scintillation bandwidth, $\Delta\nu_d$, assuming $\tau=1/2\pi\Delta\nu_d$ \citep{Rickett1977, Reardon2019}.
The scintillation bandwidth was measured by using a least-squares fit of the 2D auto-covariance function to the analytical model $C(\tau, \delta\nu)$ described in Equations 1 and 2 from \citet{Reardon2019}.
We measure the median scattering broadening time to be 5.8\,ns, the mean time to be 7.9\,ns and the standard deviation to be 6.7\,ns.
The median and the standard deviation of the diffractive scintillation time scale are measured to be 1200\,s, while the mean timescale is measured to be 1600\,s.
The measurements are uncorrelated between observing epochs, which is unsurprising as the refractive time scale ($\tau_r\approx$14\,days, assuming our inferred strength of scattering measurement) for the pulsar is less than the typical observing cadence.

\section{Discussion}\label{chapt:Discussion}


\subsection{Velocities of the IISM}\label{chapt:Velocities of the IISM}

We can use the values of $V_{\textrm{IISM}}$ to inform our model selection.
The expected mean of the plasma velocity in the IISM is $|V_{\textrm{IISM}}|=10$\,km\,s$^{-1}$ \citep{Goldreich1995}.
The priors on $|V_{\textrm{IISM}}|$ were set to be a uniform range of -200 to 200\,km\,s$^{-1}$.
If we first consider the isotropic case, all of the results for $|V_{\textrm{IISM}},\alpha|$ favoured distributions centered around $\approx$20\,km\,s$^{-1}$ (Tables \ref{tab:ourmodel_table} and \ref{tab:literaturemodel_table}).
Whereas for $|V_{\textrm{IISM}},\delta|$, we inferred a velocity between $\approx$85$-$112\,km\,s$^{-1}$, which is 8-10 times higher than the expected thermal speed.
For the plasma to move at this high velocity, it would likely need to be associated with a co-moving cloud of gas, or potentially a hot star with a small impact parameter \citep{Bignall2019}.
Objects within the IISM moving at this speed would cause shocks which could lead to further scattering.
However, we consider these scenarios to be unlikely, as we do not see any such associations, discussed further in Section \ref{chapt:Screen Associations}.


For the anisotropic models, we are sensitive to the velocity in the direction of anisotropy, $|\textbf{V}_{\textrm{IISM}},\zeta|$ varies depending on which solution for $\Omega$ is chosen.
We find that for lower velocities, a lower value for $\Omega$ is preferred.
It is possible that significant variations in $V_{\textrm{IISM}}$ would be possible over $\approx$13\,years of data.
This could be seen in our results as a step change in the arc curvature at a specific MJD if there is a sudden change in $V_{\textrm{IISM}}$.
Our data is not consistent with this scenario.
As above (Section \ref{chapt:Alternative Models}), the $\log{\rm BF}$ disfavoured test models with multiple screens with different values for $V_{\textrm{IISM}}$.
We conclude that $V_{\textrm{IISM}}$ does not vary significantly across the span of our data set.

As stated in Section \ref{chapt:Observation-Observation Parameter Variations}, we determined that each observation samples an independent sight-line through the scattering screen.
Using this we can make physical interpretations of IISM parameter variations on short timescales.
We measure this through excess variance in our data, which is modelled with the white noise parameters, $F$, and $Q$.
For $s$ we find that an observation-to-observation change of $\approx$70\,pc would be required to induce the variations we see in $f_{tn}$.
A change of this magnitude would be physically unlikely, as described in Section \ref{chapt:Observation-Observation Parameter Variations}.
Perturbations of $|V_{\textrm{IISM}}|<10$\,km\,s$^{-1}$ are expected to be the result of sampling a small part of the turbulent IISM between observations.
The variations in $|V_{\textrm{IISM}},\alpha|=5\pm1\,$km\,s$^{-1}$ are consistent with small-scale changes in the velocity of the plasma.
For $|V_{\textrm{IISM}},\delta|=15\pm3\,$km\,s$^{-1}$ we see modestly larger changes.
For the anisotropic modelling, we see changes in the $\textbf{V}_{\textrm{IISM}},\zeta=6\pm1\,$km\,s$^{-1}$ and $\zeta=3\pm1^\circ$.
Based on these inferred velocity variations, we find that the anisotropic model explains the variations between observations better than the isotropic model.

\subsection{Constraining Anisotropy}\label{chapt:Constraining Anisotropy}

We have found various arguments in favour of anisotropy in the IISM towards PSR~J1909$-$3744.
We observe this in the morphology of our arcs in the Doppler profile, and in the expected velocity of the IISM.
As there is no significant Bayesian evidence from the arc curvatures alone that supports either isotropic or anisotropic scattering, we conclude the anisotropic, if present, is likely to be weakly anisotropic ($A_r<2$).
While we are unable to precisely constrain the level of anisotropy, we estimate from comparisons with simulations that $A_r$ is not extreme, but must be at least 1.2.
As stated above, as $\psi$ changes, the morphology of scintillation arcs in the secondary spectrum also varies (Figure \ref{fig:anis_morphology}).
We see these effects in a small amount of our dataset ($<$20\%).
While we do see a broadening of the arcs power spectrum for $\approx$90\% of observations.
It is possible that this broadening is the result of unresolved arclets, which are not visible at our resolution.
This too suggests that anisotropy is present in our observations.

Therefore we have three pieces of evidence that support anisotropic scattering in our observations.
Firstly, we find more physically likely values for $|V_{\textrm{IISM}}|$ in anisotropic models.
Secondly, through comparison of simulations, we find our arc morphologies consistent with anisotropic scattering with $A_r\gtrsim1.2$.
Finally, we identify variations in arc morphology that depend on $\psi$, which is expected in anisotropic cases.
We conclude that this dataset supports a low axial ratio resulting in mild anisotropic scattering.
Highly anisotropic scattering would be manifested differently.
For instance, highly anisotropic scattering has been shown to create inverted arclets with common curvatures to the main arc, and a deep valley of power interior to the arc.
We do not observe this in our data set.
More detailed study of anisotropy would be possible with observations at higher resolutions and $S/N$ ratios.

\subsection{Orbital Dynamics}\label{chapt:Orbital Dynamics}

Scintillation can be more sensitive to measuring the inclination angle and the longitude of ascending node of binary pulsar systems.
Our initial models gave results that were inconsistent with pulsar timing.
Through model comparison, we favoured the inclination angle to be $i=86.46\pm0.05^\circ$ for this system \citep{Reardon2021}.
Based on this inclination there were two independent angles of ascending node preferred for the system, each with two measurements from recent pulsar timing analyses \citep{Liu2020, Reardon2021}.
Some models were excluded given improbably high IISM velocities, see Section \ref{chapt:Velocities of the IISM}.
As a result, the most favorable model was determined to be the anisotropic model using a $\Omega_{\textrm{prior}}=225\pm3^\circ$ \citep{Reardon2021}, giving a posterior value of $\Omega=226\substack{+4\\-3}^\circ$.
We found that the longitude of ascending node was misaligned ($30$\,$^\circ$) with the proper motion of the pulsar, found to be at an angle East of North $\theta_{PM}=196.4$\,$^\circ$.

\subsection{Screen Associations}\label{chapt:Screen Associations}

With a well-measured distance to PSR~J1909$-$3744 of $D=1158\pm3$\,pc and $s=0.49\pm0.04$ we estimate the screen distance to be $D_s=590\pm50$\,pc, and we can explore potential associations along the line of sight at this distance.
Previous works have found scattering screen associations with supernovae remnants \citep{Yao2021, Main2021}, star-forming HII regions \citep{Gupta1994, Mall2022}, the Loop I bubble \citep{Bhat1998}, the local bubble, \citep{Reardon2020, Stinebring2022} and hot stars (type O-B-A stars) \citep{Walker2017}.
We first searched the Southern H$\alpha$ Sky Survey Atlas \citep{Gaustad2001} for anomalous H$\alpha$ sources along the line of sight.
There is no evidence for unusual structures within 0.2\,degrees (corresponding to a physical scale of 2\,pc at the distance of the screen), which is not surprising given the high Galactic latitude of the pulsar.
We also searched the Gaia DR3 catalogue \citep{Gaia2018} for evidence of hot stars close to the line of sight.
We found 421 stars consistent with the screen distance and at an impact parameter of less than 2\,pc from the line of sight \citep{Walker2017}.
Of these stars, $\approx$100 had reliable temperature measurements with all stars being cooler than A-class stars, therefore not hypothesised to cause extreme scattering \citep{Walker2017}.
While other studies have found the local bubble to contribute to the scattering of pulsars \citep{Stinebring2022,McKee2022}, the screen distance excludes that being the case here.

\section{Conclusions}\label{chapt:Conclusions}

We have presented an analysis of annual and orbital variations in scintillation arc curvature for the binary PSR J1909$-$3744.
This has been accomplished across a data span of $\approx$13\,years where we have obtained 57 unique scintillation arc-curvature measurements.
We have excluded the possibility of multiple dominant scattering screens and any significant changes in the IISM parameters across the dataset.
We were able to explore a low $S/N$ regime by producing probability distributions of the power in the normalised secondary spectrum.
A comparison of the Doppler profile in our observations and simulated spectra revealed mild anisotropic scattering with $A_r\gtrsim1.2$.
Therefore, the majority of the scattering effects we see are consistent with a single, slightly anisotropic, screen.
From this, we can determine a relative screen distance approximately halfway between the pulsar and the Earth at $D_s=590\pm50$\,pc.

We have used Bayesian inference for parameter estimation and model comparison.
Using informative priors from pulsar timing, we find the data strongly supports $i=86.46\pm0.05^\circ$.
We explore our models of $|V_{\textrm{IISM}}|$, and exclude those with implausibly high values $|V_{\textrm{IISM}}|>>10$\,km\,s$^{-1}$, compared with the expected thermal speed.
This allowed us to conclude a value for the longitude of ascending node with a posterior of $\Omega=226\pm4^\circ$.
We investigated if the excess scatter in the arc-curvature measurements (quantified with white noise parameters) could be caused by astrophysical changes in the scattering screen.
It was found that these variations could be explained by small scale rms variations in $|V_{\textrm{IISM}}|\approx10$\,km\,s$^{-1}$ or the anisotropy angle $\zeta=3\pm1^\circ$.
However, the required change in $s$ was found to be physically implausible.

Further investigations will benefit from higher $S/N$ ratios and better resolution in the secondary spectrum.
This may be the focus of future work with telescopes (such as MeerKAT \citep{Bailes2020}) or receivers (such as the Murriyang ultra-wideband \citep{Hobbs2020}), that can sample a larger observational bandwidth which will translate to greater resolution in the secondary spectrum.

\section*{Acknowledgements}

We thank B. Goncharov and M. Miles for comments on the manuscript.
Murriyang, the Parkes radio telescope is part of the Australia Telescope National Facility (https://ror.org/05qajvd42) which is funded by the Australian Government for operation as a National Facility managed by CSIRO.
We acknowledge the Wiradjuri people as the traditional owners of the Observatory site.
We acknowledge use of the corner package \citep{corner}.
RMS acknowledges support through the Australian Research Council Future Fellowship FT190100155.
Part of this work was undertaken as part of the ARC Centre for Excellence for Gravitational Wave Discovery (OzGrav, CE17010004).

\section*{Data Availability}

The data and code will be made available upon reasonable request to the corresponding author.
 



\bibliographystyle{mnras}
\bibliography{J1909_arcs_Final.bbl} 




\appendix

\section{Astrophysical Sources of Jitter in Arc-Curvature Measurements}\label{chapt:Observation-Observation Variations in eta methods}

Through the use of scintillation, we can probe the IISM across long periods of time.
Our results are also sensitive to day-to-day changes in the IISM. 
This is the method outlining how we determined a relationship between small-scale changes in the IISM to variations we see from observation to observation in $f_{tn}$.
Observation-to-observation variations exist because of the independent lines of sight probed by the observations.


The scatter in an astrophysical parameter that is related to the scatter in the arc-curvature measurements.
This can be calculated using the derivative of the arc-curvature equation (Equation \ref{eq:fdop}) with respect to the parameter.

Therefore, we can quantify changes in the IISM by taking the derivative of the following parameters with respect to $f_{tn}$; $s$, $V_{\textrm{IISM},\alpha}$, $V_{\textrm{IISM},\delta}$, $\zeta$ and $V_{\textrm{IISM},\zeta}$.

First we re-arrange Equation \ref{eqn:eta} in terms of $f_{tn}$, 
\begin{eqnarray}
    f_{tn} = |\textbf{V}_{\textrm{eff}}(s)|\cos(\psi)\sqrt{\frac{2\eta_0}{Ds(1-s)}},
	\label{eq:fdop}
\end{eqnarray}
and define the expanded isotropic and anisotropic equations, respectively, for $V_{\textrm{eff}}(s)$ from Equation \ref{eq:veff} seen in Equations \ref{eq:isot_veff} and \ref{eq:anis_veff}.
We include the overall change in $s$ and $V_{\textrm{eff}}(s)$ for both isotropic and anisotropic modelling,
\begin{eqnarray}
    \frac{\delta f_{tn}}{\delta \textbf{V}_{\textrm{eff}}(s)} = \cos{\psi}\sqrt{\frac{2\eta_0}{Ds(1-s)}},
	\label{eq:derivitive_veff}
\end{eqnarray}
\begin{eqnarray}
    \frac{\delta f_{tn}}{\delta s} = \frac{V_{\textrm{eff}}(s)\cos{\psi}\left(\eta_0 / \left( D\left(1 - s\right)^2s \right) - \eta_0 / \left( D\left(1 - s\right)s^2 \right) \right)}{\sqrt{ \left( 2\eta_0\right) / \left( Ds\left(1 - s\right) \right)}},
	\label{eq:anis_derivitive_s}
\end{eqnarray}
where for the isotropic case $\cos{\psi}=1$.
For the isotropic model using Equation \ref{eq:isot_veff} for $V_{\textrm{eff}}(s)$, we determined the following equations,
\begin{eqnarray}
    \frac{\delta f_{tn}}{\delta V_{\textrm{IISM},\alpha}} &&= -\sqrt{ \left( 2\eta_0\right) / \left( Ds\left(1 - s\right) \right)} \nonumber\ \\
    && \times \frac{\left(V_{\textrm{kin},\alpha} - V_{\textrm{IISM},\alpha}\right)}{V_{\textrm{eff}}(s)},
	\label{eq:derivitive_vism_ra}
\end{eqnarray}
\begin{eqnarray}
    \frac{\delta f_{tn}}{\delta V_{\textrm{IISM},\delta}} &&= -\sqrt{ \left( 2\eta_0\right) / \left( Ds\left(1 - s\right) \right)} \nonumber \\
    && \times \frac{\left(V_{\textrm{kin},\delta} - V_{\textrm{IISM},\delta}\right)}{V_{\textrm{eff}}(s)}.
	\label{eq:derivitive_vism_dec}
\end{eqnarray}
For the anisotropic model using Equation \ref{eq:anis_veff} for $V_{\textrm{eff}}(s)$, we determined the following equations,
\begin{eqnarray}
    \frac{\delta f_{tn}}{\delta \zeta} &&= \sqrt{ \left( 2\eta_0\right) / \left( Ds\left(1 - s\right) \right)}
    \nonumber \\ 
    && \times \left(V_{\textrm{kin},\alpha}\cos{\zeta} - V_{\textrm{kin},\delta}\sin{\zeta}\right)
    \label{eq:derivitive_psi}
\end{eqnarray}
\begin{eqnarray}
    \frac{\delta f_{tn}}{\delta V_{\textrm{IISM}_{\zeta}}} &&= \sqrt{ \left( 2\eta_0\right) / \left( Ds\left(1 - s\right) \right)} \nonumber\ \\
    && \times \frac{\left(V_{\textrm{kin},\alpha}\sin{\zeta} + V_{\textrm{kin},\delta}\cos{\zeta}V_{\textrm{IISM}_{\zeta}}\right)}{V_{\textrm{eff}}(s)}.
    \label{eq:derivitive_vism_psi}
\end{eqnarray}

Using these equations we can input the measured parameters from modelling.
Then we evaluate the change necessary in that parameter using the rms of $f_{tn}$.
We do this for both the isotropic and anisotropic models separately.
Our results and discussion on this are presented in Section \ref{chapt:Observation-Observation Parameter Variations} and \ref{chapt:Velocities of the IISM} respectively.
We estimated the errors on these parameters assuming sampling errors.
Such that the error in a parameter was given by 
\begin{eqnarray}
    \sigma_P = \sigma_t P,
    \label{eq:parameter_error}
\end{eqnarray}
where $P$ is the parameter estimation,
\begin{eqnarray}
    \sigma_t = \left(\frac{2 t^4}{N(N-1)}\right)^{1/4},
    \label{eq:parameter_error_extra}
\end{eqnarray}
and $t$ is the rms of $f_{tn}$.

It is possible that we are biased against detecting high-curvature arcs as we would be less sensitive to these because of both noise at the $f_t=0$ axis and low resolution in the secondary spectra.
We compared our arc curvature measurements with the distribution of expected measurement across this MJD space (assuming observations have been scheduled independent of arc curvature measurements, which they have).
Figure \ref{fig:KolmogorovSmirnov} shows the cumulative distribution functions of both.
A Kolmogorov-Smirnov test comparing the returns a p-value of 0.03-0.04 for the isotropic and anisotropic model, respectively.
This suggests with modest confidence that we are missing high arc curvature measurements.
This is shown in Figure \ref{fig:KolmogorovSmirnov}, which used the anisotropic model from Figure \ref{fig:model_anis}.

\begin{figure}
	\includegraphics[width=1\columnwidth]{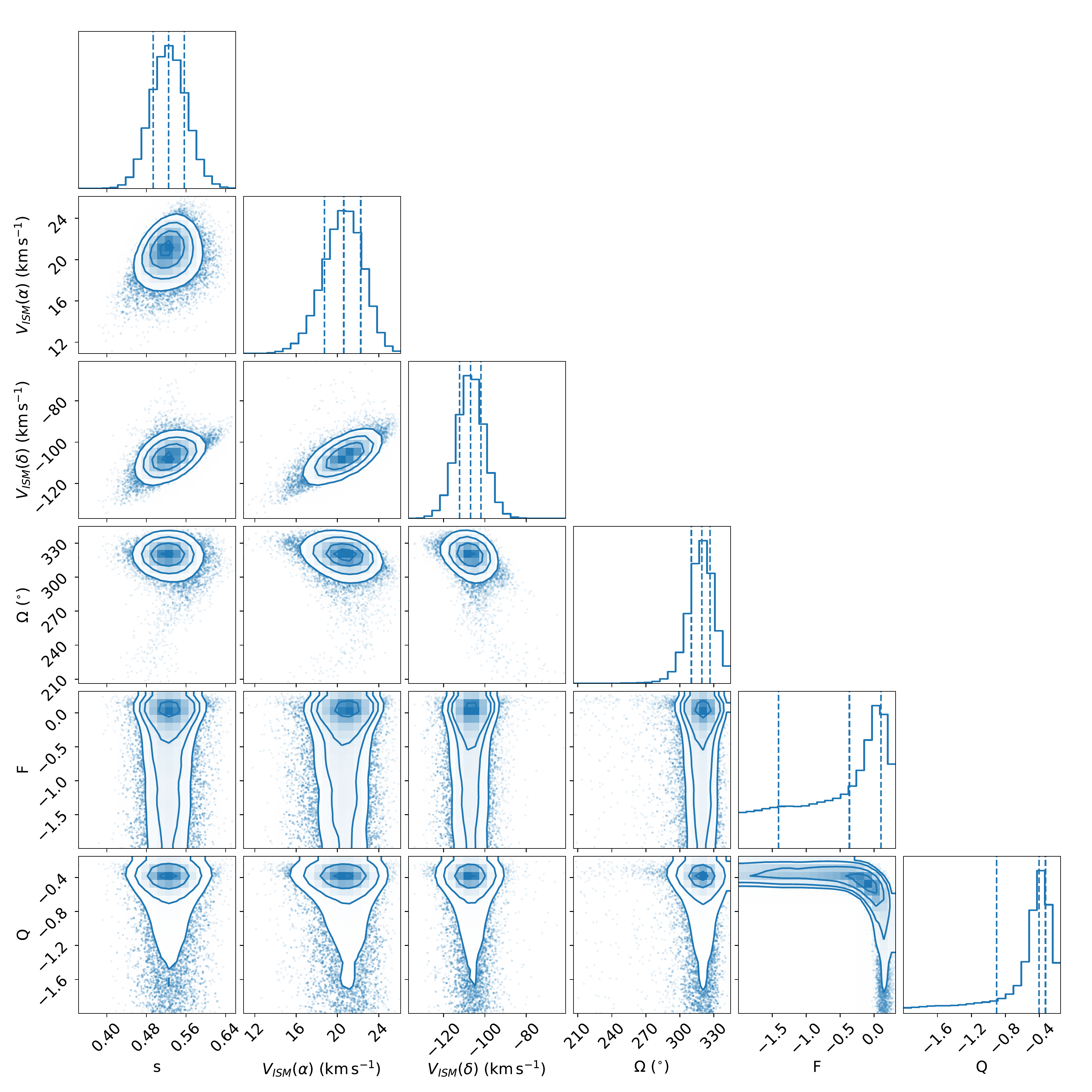}
    \caption{ One and two-dimensional marginal posterior probability distributions, assuming isotropic scattering.}
    \label{fig:corner_isot}
\end{figure}

\begin{figure}
	\includegraphics[width=1\columnwidth]{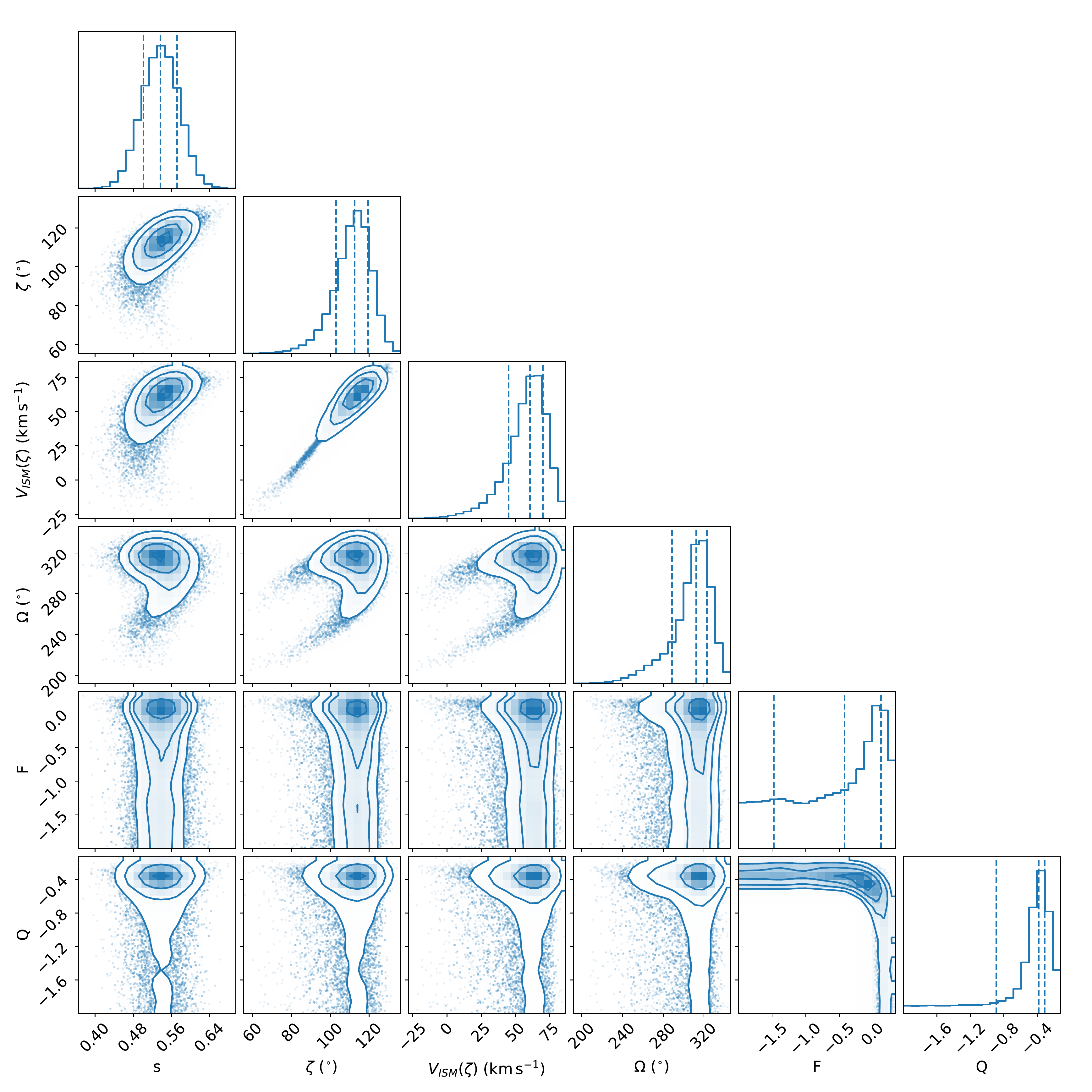}
    \caption{ One and two-dimensional marginal posterior probability distributions, assuming anisotropic scattering.}
    \label{fig:corner_anis}
\end{figure}

\begin{figure}
	\includegraphics[width=1\columnwidth]{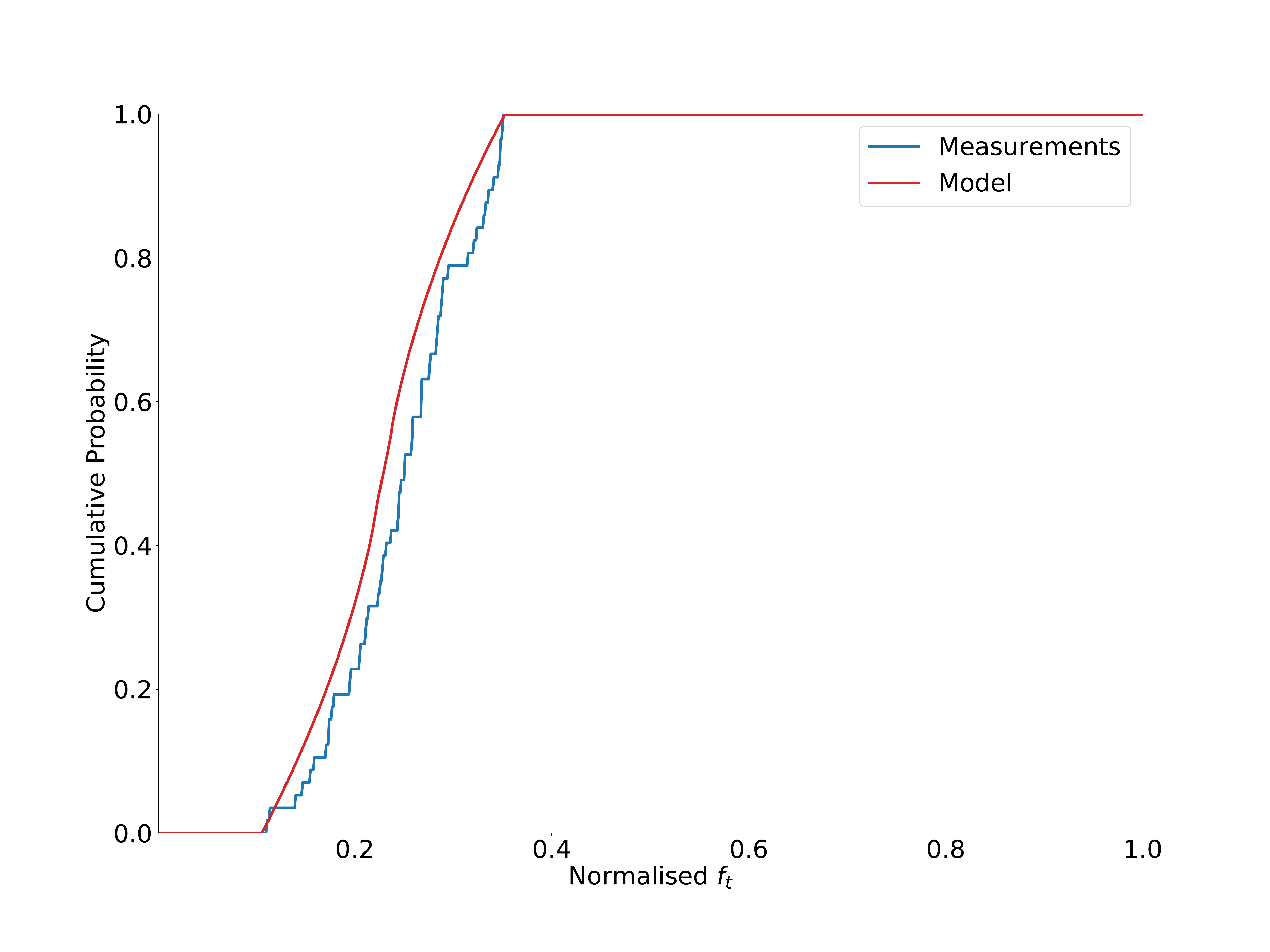}
    \caption{ Cumulative distribution of normalised arc curvature measurements (blue) and model-predicted values (red). A Kolmogorov-Smirnov test shows we are biased against detecting high arc curvatures, which is equivalent to lower values of $f_{tn}$. We attribute this to RFI-induced noise at $f_t$=0 and poor resolution.}
    \label{fig:KolmogorovSmirnov}
\end{figure}

\begin{figure}
	\includegraphics[width=1\columnwidth]{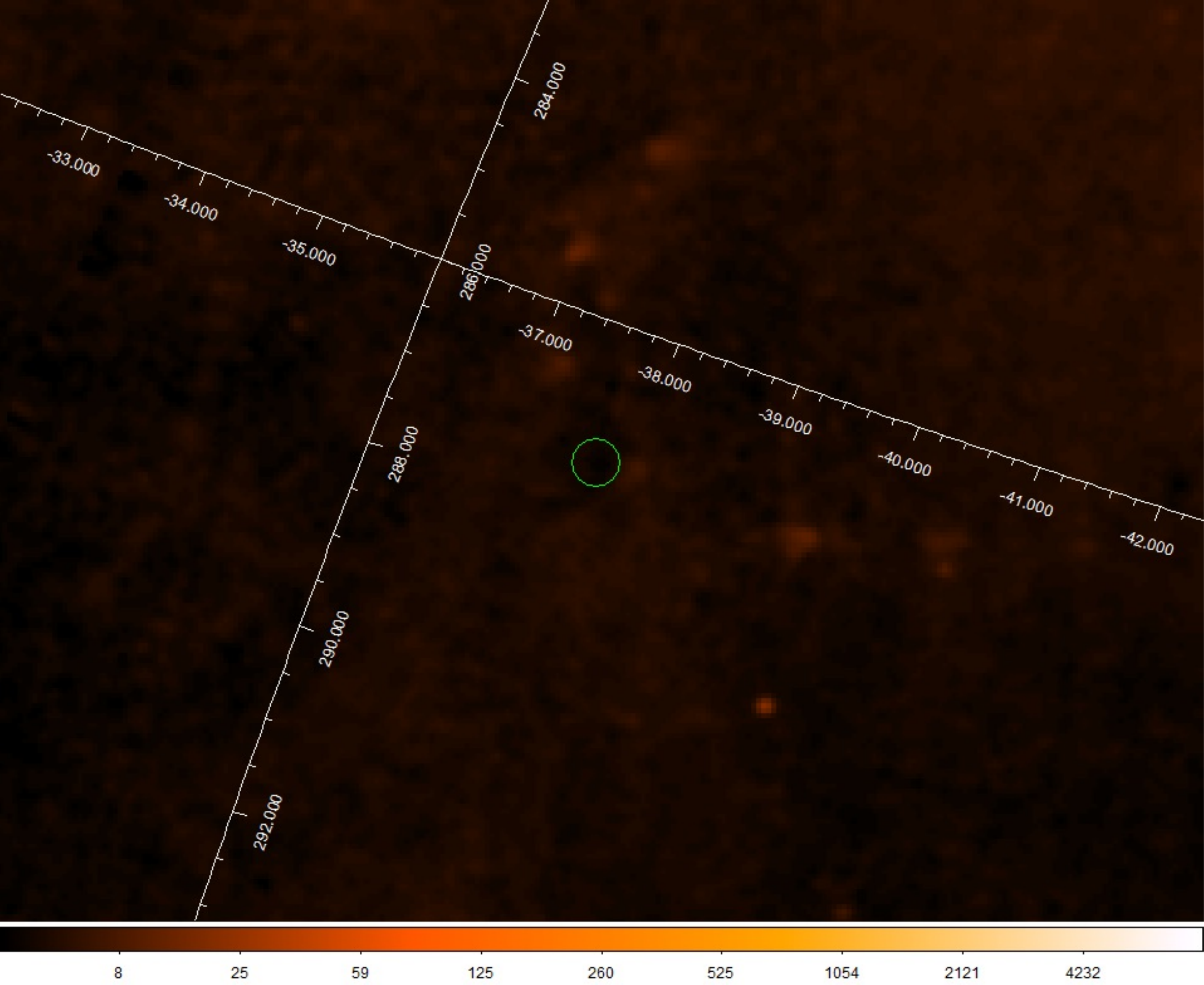}
    \caption{ H$\alpha$ map in the direction of J1909$-$3744. This map was produced from the Southern H$\alpha$ Sky Survey Atlas \citep{Gaustad2001}, where the colour bar represents the intensity in units of ergs\,cm$^{-2}$\,sec$^{-1}$. The green region represents a 0.2$^{\circ}$ angle caused by the impact parameter of 2\,pc at the estimated screen distance $D_s=590\pm50$\,pc.}
    \label{fig:Halpha}
\end{figure}




\bsp	
\label{lastpage}
\end{document}